# Two-Layered Superposition of Broadcast/Multicast and Unicast Signals in Multiuser OFDMA Systems

David Vargas and Yong Jin Daniel Kim, *Member*, IEEE

*Abstract*— We study optimal delivery strategies of one common and $K$ independent messages from a source to multiple users in wireless environments. In particular, two-layered superposition of broadcast/multicast and unicast signals is considered in a downlink multiuser OFDMA system. In the literature and industry, the two-layer superposition is often considered as a pragmatic approach to make a compromise between the simple but suboptimal orthogonal multiplexing (OM) and the optimal but complex fully-layered non-orthogonal multiplexing. In this work, we show that only two-layers are necessary to achieve the maximum sum-rate when the common message has higher priority than the $K$ individual unicast messages, and OM cannot be sum-rate optimal in general. We develop an algorithm that finds the optimal power allocation over the two-layers and across the OFDMA radio resources in static channels and a class of fading channels. Two main use-cases are considered: i) Multicast and unicast multiplexing when $K$ users with uplink capabilities request both common and independent messages, and ii) broadcast and unicast multiplexing when the common message targets receive-only devices and $K$ users with uplink capabilities additionally request independent messages. Finally, we develop a transceiver design for broadcast/multicast and unicast superposition transmission based on LTE-A-Pro physical layer and show with numerical evaluations in mobile environments with multipath propagation that the capacity improvements can be translated into significant practical performance gains compared to the orthogonal schemes in the 3GPP specifications. The impact of real channel estimation is also analyzed and it is shown that significant gains in terms of spectral efficiency or increased coverage area are still available even in the presence of estimation errors and imperfect interference cancellation for the two-layered superposition system.

## I. INTRODUCTION

THE landscape of media consumption is radically changing thanks to the proliferation of smartphones and tablets with mobile internet connectivity. Users can now experience media services in different environments, from in-home to on the move where the consumption of live or on-demand TV programmes can be combined with the interaction of social media networks. Furthermore, the Media and Entertainment industry is evolving towards media experiences that are increasingly personal, interactive and immersive where the audio-visual content is changed according to the user requirements. *Object-based Media*, where the creation and storage of the elements of a film, Radio or TV programme is done as media *objects*, allows content to be rendered at the receivers in the form most suitable for a particular user (e.g. additional camera angles of an live sport event or changing the presenter for hearing-impaired viewers) [1]-[2].

The efficient wireless delivery of these existing and new experiences to multiple users require mobile broadband technologies with the capability to multiplex into the same system broadcast/multicast signals (to convey common information, e.g., live TV) and unicast signals (to convey independent messages for personalisation). In this paper, unicast refers to one-to-one communication between a transmitter and a receiver with uplink capability that can feed back the channel-state-information (CSI). Both multicast and broadcast refer to the transmission of a common message to multiple receivers, but in multicast the receivers have uplink capabilities and can feedback to the transmitter CSI whereas in broadcast the receivers lack uplink capabilities and/or cannot feedback the CSI. The 3rd Generation Partnership Project (3GPP) introduced Multimedia Broadcast Multicast Services (MBMS) in the third and fourth generation mobile broadband specifications (3G and 4G), where broadcast/multicast and unicast signals are allocated to separate time and frequency resources, i.e., orthogonal multiplexing (OM), cf. Fig. 1a. However, with the development of the fifth generation (5G) mobile broadband standards, there is an increasing interest in the industry in non-orthogonal multiplexing (NOM) approaches that can increase the spectral efficiency (SE) by transmitting superimposed signals with different robustness requirements. In this paper, we study optimal delivery strategies of common and independent messages from one source to multiple receivers in wireless environments [1]. Particularly, we consider a two-layered superposition of broadcast/multicast and unicast signals in the downlink multiuser scenario setup where different unicast signals are allocated orthogonal resources, cf. Fig. 1b.

### A. Related Work

The theoretical analysis of the single-input single-output Gaussian downlink multiuser channel with $K$ independent plus

D. Vargas is with the BBC Research and Development, The Lighthouse, White City Place, 201 Wood Lane, London, W12 7TQ, U.K., (email: david.vargas@bbc.co.uk).
Y.J.D. Kim is with the Department of Electrical and Computer Engineering, Rose-Hulman Institute of Technology, Terre Haute, IN 47803 USA (e-mail: kim2@rose-hulman.edu).

---

[1] The term broadcast channel, as used in information theory, often refers to unicast transmission in downlink channel. In this paper, we give a broader meaning to the term and use it to refer to a channel where a single source communicates any types of information simultaneously to multiple receivers.

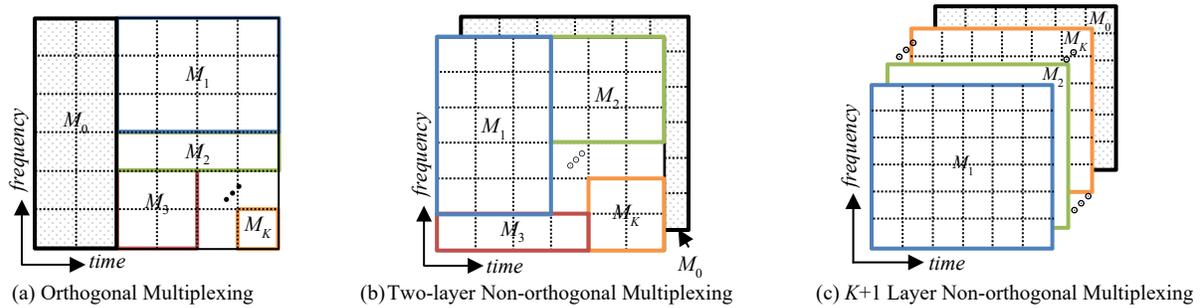

Fig. 1. Comparison of different multiplexing methods on OFDMA: a) the conventional orthogonal multiplexing, b) the considered two-layer non-orthogonal multiplexing, and c) $K+1$ layer non-orthogonal multiplexing.

one common messages to $K$ receivers began in the 1970's, showing that the capacity is achieved by NOM of all messages in a $K+1$ layer architecture, cf. Fig. 1c, followed by multiuser decoding at each receiver [3]. Despite this seminal work, more than 30 years later the physical layer of 4G Long Term Evolution - Advanced Pro (LTE-A-Pro) [4] and the recently standardised 5G New Radio (NR) physical layer [5] use Orthogonal Frequency Division Multiple Access (OFDMA) as the principal multiple access technology, motivated by the reduced receiver complexity and flexible implementation of user scheduling algorithms. OFDMA assigns orthogonal time and frequency resources to signals, and in LTE-A-Pro the multiplexing of broadcast/multicast and unicast signals can only be realized orthogonally by evolved MBMS (eMBMS). Without a common message, the capacity region of downlink multiuser OFDMA with $K$ unicast messages along with the optimal power allocation are characterized in [6]. It is shown that the maximal sum-rate is achieved by OM where one user with the best channel gain is assigned to each subcarrier and the power is distributed across the subcarriers by a water-filling policy [6]-[7]. In the presence of a common message, the capacity region and the optimal power allocation are characterized for the case of two parallel channels in [8] and for more than two channels in [9], but for both references the scope was limited to two users only. In the case of two-user multiple-input-multiple-output (MIMO) downlink channel with a common message, the capacity region is achieved by superposition coding combined with vector dirty paper coding (DPC) [10]-[11]. Due to the non-degradedness nature of MIMO downlink channels, however, NOM based on the superposition coding principle without DPC, as considered in this work, becomes strictly suboptimal in MIMO.

NOM of unicast signals has attracted much attention from the research community [12]-[18]. Also known as multiuser superposition transmission (MUST) [19], NOM has been standarised in 3GPP for LTE-A-Pro for the superposition of two unicast signals, but broadcast or multicast is not supported. On the other hand, NOM of two broadcast signals, also known as layered division multiplexing (LDM) [20]-[22], has been included in the digital terrestrial broadcast standard ATSC 3.0 targeting robust mobile reception (with low Signal-to-Noise Ratio (SNR) target) and delivering Ultra High Definition TV (UHDTV) to fixed rooftop antennas (high SNR target). Detailed capacity-coverage performance analysis of LDM [23] shows that it can outperform OM-based approaches in many practical scenarios. LDM, in its original form, targets two different types of terminals where each type receives a common message, and is generally different from the unicast transmission where each user requests independent message.

The topic of joint broadcast/multicast and unicast transmission is starting to get more attention in the recent research literature. In [25]-[27], the superposition of a multicast and a unicast message is considered with the users employing successive interference cancellation (SIC) to decode and remove the multicast message before decoding the unicast message. However, the scope is limited to transmission of a single unicast message targeting either one particular user in the network [27] or the user with the highest channel quality [25]-[26]. Transmission of more than one unicast messages is considered in [28]-[31]. In these works, after decoding and subtracting the multicast signal, each user decodes the desired unicast signal while treating unicast signals that are intended for other users as additional noise. In order to minimize this interference between the unicast signals, linear beamforming in the spatial domain is proposed in [28] as well as cooperation between the base-stations in a multi-cell downlink in [29]-[31]. In [32], the broadcast message is placed in the null-space of the channel collectively seen by the unicast users so that the broadcast message does not interfere with the unicast. A different technique based on rate-splitting (RS) is proposed in [33]-[34], in which dynamically chosen proportions of the unicast messages are integrated into the multicast message so that they too may be decoded by all users – thus allowing further reduction of the interferences seen by each user and improving the performance compared to the conventional linear beamforming or the (power domain) NOM systems. [34] considered both 1-layer and multi-layer RS transmissions and showed that the 1-layer RS with a single stage of SIC is able to capture most of the performance benefits of the generalized multi-layer RS at much reduced complexity.

In the aforementioned works [28]-[34], multiple unicast messages are non-orthogonally multiplexed over all time and frequency. Alternatively, the unicast messages may be *orthogonally* multiplexed over OFDM resources and broadcast message may be superimposed on top. This two-layer NOM over OFDM is shown to have performance benefits over OM through system level simulations in [24]. The focus of the study, however, was limited to two layers to minimize the impact of error propagation without theoretical justification on its optimality or optimization of resource allocation over OFDM channels. Within the standardisation bodies, superposition of broadcast/multicast and unicast signals has not been considered yet.

### B. Contributions of this Paper

In this work, we consider two-layered superposition of broadcast/multicast and unicast signals in two use-cases called multicast-unicast (MU) and broadcast-unicast (BU) multiplexing. In the MU multiplexing use-case, a base station communicates a common message (e.g. sport event) as well as $K$ independent messages (for personalisation) to $K$ user terminals with uplink capabilities. In the BU multiplexing use-case, a base station transmits a common message to a receiving population without uplink capabilities and additionally transmits $K$ independent messages (as well as the common message) to $K$ users with uplink capabilities. In both use-cases, we assume that the common message (broadcast/multicast signal) has higher priority than any of the individual unicast messages[2]. We note that the two-use cases are often used interchangeably in both the research literature and industry but have important differences that affect the system design.

As seen in the related work section, most of the prior research literature and industry efforts focus on the NOM of either unicast signals only (e.g. MUST in 3GPP) or broadcast/multicast signals only (e.g. LDM in ATSC 3.0). While there is a growing interest on the NOM of broadcast/multicast and unicast signals from the research community, detailed system modeling and analysis of the corresponding rate regions in the multiuser downlink channel with OFDMA is missing (e.g., the system model is either limited to single-carrier, a common message with a single unicast message, or two-users). Moreover, there has not yet been a theoretical analysis on how far the two-layer NOM approach can perform from the $K$-user downlink multiuser channel capacity. In addition, most of the works in the research literature on the two-layer superposition of broadcast/multicast and unicast signals have either provided a theoretical analysis or a performance assessment from the practical point of view but do not address both aspects in a unified manner.

The contributions of this paper are summarized next:
- *We prove sum-rate optimality of two-layer superposition of broadcast/multicast and unicast signals in downlink multiuser OFDMA systems in static and a class of fading channels. The main implication is that achieving the optimal weighted sum-rates needs more than one layer but no more than two layers.* This result generalizes [9] to more than two users and non-static channels. Unlike [9], where OM was observed to be performing close to the capacity of NOM, we show that with more than two users, the multiuser diversity for unicast signals and multicast gain of broadcast/multicast signals can provide significant gains over OM.

- *We develop an optimal power allocation algorithm over the two-layers and across the OFDMA radio resources.* The algorithm is used to solve two constrained sum-rate maximization problems in the MU and BU use-cases. The algorithm reveals significant capacity potential of the two-layer superposition as compared to OM, and the capacity gain generally increases with increasing number of users.

- *The two-layer architecture is sum-rate capacity optimal, enhances cell-edge user performance and retains decoding simplicity at the receivers.* The considered two-layer architecture achieves the same capacity point as the $K+1$ NOM approach (cf. Fig. 1c) and retains the simplicity of a single stage of interference cancellation with each receiver decoding only wanted information. Additionally, since the rate of the common message is increased and decoded by all the users, the considered two-layer architecture also enhances the cell-edge user performance.

- *Finally, we present the two-layered broadcast/multicast and unicast superposition transceiver architecture based on LTE-A-Pro physical layer.* The simulation results in realistic mobile environments with challenging multipath propagation verify that the NOM capacity improvements can be translated onto significant practical performance gains compared to the OM schemes in the 3GPP specifications. Our results also incorporate the impact of realistic channel estimation and imperfect interference cancellation into the performance of the two-layer system. Our results show similar degradations for our two-layered transceiver architecture and OM, which confirms that significant performance gains are still available with practical receiver implementations.

## II. SYSTEM MODEL AND SCOPE OF THE STUDY

In this work, we consider a single-cell downlink transmission of broadcast/multicast and unicast contents over OFDM resources. A broadcasting transmitter sends a common message $M_0$ at rate $R_0$ and $K$ independent messages $M_1, …, M_K$ at rates $R_1, …, R_K$, respectively. If a certain receiver wants only the common message, the corresponding user's independent message rate can be set to zero. Each time/frequency slot in an OFDM grid is referred to as a resource element (RE) and the total number of REs available to transmit the $K+1$ messages is denoted by $N$. Without loss of generality, we divide OFDM resources into $M$ "parallel" channels, each consisting of $N^{(1)}$, $N^{(2)}, …, N^{(M)}$ REs respectively, so that $\sum_{i=1}^{M} N^{(i)} = N$. The downlink scheduler in radio resource management can arbitrarily allocate user messages to these parallel channels based on different optimization criteria. We allow allocating more than one messages to a channel using NOM.

We assume that the delay spread of multipath channel is small compared to the symbol time of individual subcarriers of OFDM so that each subcarrier experiences flat fading and the impact of intersymbol interference (ISI) can be mitigated via cyclic prefix. Then, $n^{th}$ received signal of the $k^{th}$ receiver in the $i^{th}$ parallel channel may be expressed as

$$Y_k^{(i)}[n] = H_k^{(i)}[n] X^{(i)}[n] + W_k^{(i)}[n], \quad (1)$$

where $H_k^{(i)}[n]$ denotes a flat fading channel gain between the transmitter and $k^{\text{th}}$ receiver in the $i^{\text{th}}$ channel, $X^{(i)}[n]$ is the $n^{\text{th}}$ complex modulated symbol, and $W_k^{(i)}[n]$ is a zero mean Gaussian noise with variance $\sigma_k^2$.[3]

---

[2] This is a reasonable assumption for the two use-cases considered. In the MU multiplexing, the common message needs to be delivered to the $K$ users that requested the content and its data rate contributes to rate experienced by each user individually. As for BU multiplexing, the common message is the primary service while unicast signals can use the spectrum resources with the restriction of guaranteeing certain quality of service to the common message (e.g. coverage or data rate).

We consider the static channels and a class of fading channels in our analysis. In the static channels, $H_k^{(i)}[n]$ are assumed constant over $n$ (but not necessarily over $i$ or $k$) and the transmitter is assumed to know the full CSI, i.e. instantaneous amplitude and phase of the channel realisations. The static channel model characterizes the case when users are relatively stationary over the duration of an OFDM block. In the fading channel, $H_k^{(i)}[n]$ are assumed randomly varying over $n$, assuming that the code block length is long enough to span many channel coherence time intervals and to average over multiple fading states of the channel. In the fading case we assume partial CSI at the transmitter, i.e., knowledge of channel statistics and long-term averaged SNR values of each user[4]. Due to the inability to track the time-varying fading channel gains, the power assigned to each channel is fixed over the time duration of the channel but different power can be assigned to different channels and to different user messages. In all cases, we assume that the receivers can estimate their own CSIs.

### III. Capacity Regions and Optimality of Two-Layer Superposition in Static and a Class of Fading Channels

In this section, we present the capacity regions of the considered channel models and characterize the cases when the two-layer superposition can be sum-rate optimal in static channels and a class of fading channels.

#### A. Capacity Region in Static Channels and Cases Where Two-Layer Superposition is Optimal

In the static channels, we may find a natural ordering of users based on their channel qualities. The effective noise power experience by the $k^{\text{th}}$ user in the $i^{\text{th}}$ channel is defined as

$$Z_k^{(i)} \equiv \tfrac{N^{(i)}}{N} \sigma_k^2 \big/ \big| H_k^{(i)} \big|^2, \quad (2)$$

where the symbol index $n$ is dropped from $H_k^{(i)}[n]$ due to invariance of the channel gains in the static channels. Let $\pi^{(i)}$ be the permutation such that $Z_{\pi^{(i)}(1)}^{(i)} < Z_{\pi^{(i)}(2)}^{(i)} < \cdots < Z_{\pi^{(i)}(K)}^{(i)}$, i.e., $\pi^{(i)}(1)$ is the index of the user that has lowest noise power, and $\pi^{(i)}(2)$ is the user with the second lowest, and so on, in the $i^{\text{th}}$ channel[5]. We define two auxiliary rate functions by:

$$R_{0k}^{(i)} \equiv \frac{N^{(i)}}{N} \log_2 \left( 1 + \frac{P_0^{(i)}}{Z_k^{(i)} + \sum_{j=1}^K P_j^{(i)}} \right) \text{ and}$$

$$R_k^{(i)} \equiv \frac{N^{(i)}}{N} \log_2 \left( 1 + \frac{P_k^{(i)}}{Z_k^{(i)} + \sum_{j \in \mathcal{G}_k^{(i)}} P_j^{(i)}} \right),$$

which respectively represent the rate of the common message component and the rate of the independent message sent to $k^{\text{th}}$ user in $i^{\text{th}}$ channel. $\mathcal{G}_k^{(i)}$ denotes the set of users whose noise power in $i^{\text{th}}$ channel are lower than that of the $k^{\text{th}}$ user, i.e., $\mathcal{G}_k^{(i)} \equiv \{ j \in [1:K] : Z_j^{(i)} < Z_k^{(i)} \}$. $P_0^{(i)}$ and $P_k^{(i)}$ denote the power assigned to the common and $k^{\text{th}}$ independent messages, respectively, and the total power allocated to the $i^{\text{th}}$ channel is given by $P^{(i)} \equiv P_0^{(i)} + \sum_{k=1}^K P_k^{(i)}$. The scaling factor $\tfrac{N^{(i)}}{N}$ is a penalty term for using only $N^{(i)}$ REs out of $N$ REs in the $i^{\text{th}}$ channel.

The capacity region of $M$ parallel Gaussian broadcast channel (GBC) with a common message and two independent messages for two users is characterized in [9]. Straightforward extension to $K$ users yields the following capacity region:

$$\mathcal{C} \equiv \bigcup_{\boldsymbol{p} \in \mathcal{P}} \left\{ \boldsymbol{R} : R_0 \le \min_{k \in [1:K]} \left\{ \sum_{i=1}^M R_{0k}^{(i)} \right\}, R_k \le \sum_{i=1}^M R_k^{(i)}; k \in [1:K] \right\}, \quad (3)$$

where $\boldsymbol{R} \equiv [R_0, R_1, \cdots, R_K]$ and the union is taken over all admissible power allocation vectors $\boldsymbol{p}$ that satisfy the following power constraint:

$$\mathcal{P} \equiv \left\{ \boldsymbol{p} : \sum_{i=1}^M P^{(i)} = \sum_{i=1}^M \left( P_0^{(i)} + \sum_{k=1}^K P_k^{(i)} \right) \le P \right\}. \quad (4)$$

This capacity region leverages the *stochastic degradedness* of GBC; i.e., if a certain user can decode a message, another user with a lower effective noise power can also decode that message. The common message is sent at a rate equal to the minimum of the individual user capacities, so that it can be recovered at *every* user while treating independent messages as noise. The independent messages are sent at rates such that they can be recovered through successive cancellations, which involves sequentially decoding and stripping off the common message and all independent messages of the users with higher noise powers than the given user.

---

[3] While our study focuses on single-antenna systems, our system model can be readily extended to multiple-antenna transmitters using diversity schemes that are transparent to the receivers (e.g., cyclic-delay diversity schemes as specified in [35]) and/or multi-antenna receivers using diversity combiners (e.g. maximum ratio combiner). The resulting composite channel is same as (1) with modified channel gains and noise variances, and the main results of our paper apply to such multi-antenna systems.

[4] In practical systems, such as 3GPP LTE, the long-term averaged SNR of each user can be derived from the channel quality indicator (CQI) feedback.

[5] In the literature, the users are often indexed, without loss of generality, by the decreasing order of the individual link SNRs. Such user indexing is possible in degraded channels such as Gaussian BC. However, because user orderings may be different in different OFDM subchannels, we do not use such indexing in this work. We further assume that no two users have exactly the same effective noise power, which does not lose generality if the probability of such event is zero (such is the case when the channel gains are independently drawn from a continuous distribution).

The capacity region $\mathcal{C}$ is convex (which may be shown using the standard time-sharing argument, see e.g., Section 14.3.3 in [3]). Thus, for every rate vector $\boldsymbol{R}$ on its boundary surface, there exist nonnegative real numbers $\mu_0$, $\mu_1$, ..., $\mu_K$ such that $\boldsymbol{R}$ is the solution to the following weighted sum-rate maximization problem:

$$\max_{\boldsymbol{R} \in \mathcal{C}} \mu_0 R_0 + \mu_1 R_1 + \cdots + \mu_K R_K. \quad (5)$$

The solution to (5) for a fixed set of $\{\mu_k\}$ results in a single boundary point on $\mathcal{C}$. All points on the boundary surface of $\mathcal{C}$ can be obtained by varying $\mu_k$'s. These numbers may also be interpreted as *rate rewards* [6] since increasing any one can give higher priority to the corresponding message (e.g., higher $\mu_1$ leads to a boundary point with higher $R_1$). Using the method of Lagrange multipliers, (5) is equivalent to

$$\max_{\boldsymbol{p} \in \mathcal{P}} \left\{ \mu_0 \min_{k \in [1:K]} \left\{ \sum_{i=1}^{M} R_{0k}^{(i)} \right\} + \sum_{k=1}^{K} \mu_k R_k - \lambda \sum_{i=1}^{M} P^{(i)} \right\}, \quad (6)$$

where $\lambda$ is a nonnegative Lagrange multiplier and $P^{(i)}$ is the total power allocated to $i^{\text{th}}$ channel.

The following proposition characterizes the cases when the one-layer (i.e., OM) or the two-layer superposition can achieve the solution to (6). This extends the results in [9] for two users to arbitrary $K$ users.

*Proposition 1 (Optimality of one-layer or two-layer superposition):* Suppose same rate rewards are given to all independent messages, i.e., $\mu_1 = \mu_2 = \cdots = \mu_K = \mu$. If $\mu_0 \leq \mu$, then in each of the $M$ parallel channels, sending a single independent message of the user with the lowest effective noise power can achieve the maximum weighted sum-rate (5). If $\mu_0 > \mu$, then in each of the $M$ parallel channels, sending two-layer superposition of the common message *and* a single independent message of the user with the lowest effective noise power can achieve (5).

*Proof:* See Section III.B. ∎

Consequently, achieving the maximum weighted sum-rate $\mu_0 R_0 + \mu(R_1 + R_2 + \cdots + R_K)$ for any $\mu_0 > \mu$ needs more than one layer (i.e., it needs some degrees of non-orthogonality) but no more than 2 layers. When the two-layer superposition is used, the capacity region $\mathcal{C}$ becomes the following rate region:

$$\mathcal{C}_{\text{2-layer}} \equiv \bigcup_{\boldsymbol{p} \in \mathcal{P}} \left\{ \boldsymbol{R} : R_0 \leq \min_{k \in [1:K]} \left\{ \sum_{i=1}^{M} \frac{N^{(i)}}{N} \log_2 \left( 1 + \frac{P_0^{(i)}}{Z_k^{(i)} + P_{\pi^{(i)}(1)}^{(i)}} \right) \right\},$$

$$R_k \leq \sum_{i=1}^{M} \frac{N^{(i)}}{N} \log_2 \left( 1 + \frac{P_k^{(i)}}{Z_k^{(i)}} \right) \mathbf{1}\left(k = \pi^{(i)}(1)\right); k \in [1:K] \right\}, \quad (7)$$

where $\mathbf{1}(\cdot)$ denotes the indicator function (i.e., $\mathbf{1}(A) = 1$ if $A$ is true and 0 otherwise) and $\pi^{(i)}(1) \in [1:K]$ is index of the user with the lowest effective noise power in the $i^{\text{th}}$ channel.

As a remark, the common message parts must be jointly encoded and jointly decoded across $M$ parallel channels, while the independent messages can be separately encoded from the common message parts and among themselves. It can be shown that if the common message components are separately encoded, the bound on $R_0$ in $\mathcal{C}_{\text{2-layer}}$ changes to a sum of the minimum, instead of the minimum of the sum. The resulting achievable rate region is in general strictly contained in $\mathcal{C}_{\text{2-layer}}$. It is interesting to note that these two rate regions coincide when the user orderings are the same across all $M$ parallel channels.

### B. Proof of Proposition 1 using the Utility Functions Method

The proof generally follows ideas outlined in [6] and [9] with important differences that here we have both a common message *and* an arbitrary number of independent messages. We start by recognizing that (6) is a *max-min* problem of the following form:

$$\max_{\boldsymbol{p} \in \mathcal{P}} \left\{ \min \left\{ G_1(\boldsymbol{p}), G_2(\boldsymbol{p}), \cdots, G_n(\boldsymbol{p}) \right\} + G_{n+1}(\boldsymbol{p}) \right\}, \quad (8)$$

where $G_1(\boldsymbol{p})$, ..., $G_{n+1}(\boldsymbol{p})$ are real continuous functions of $\boldsymbol{p}$. This class of max-min problem may be solved by a technique in [Proposition 1, [36]], as summarized in the following lemma.

*Lemma 1 [36]:* Let $\boldsymbol{\alpha} \equiv [\alpha_1, \alpha_2, \cdots, \alpha_n]$ with nonnegative $\alpha_i$'s and $\sum_{i=1}^{n} \alpha_i = 1$. Furthermore, let

$$G(\boldsymbol{\alpha}, \boldsymbol{p}) \equiv \sum_{i=1}^{n} \alpha_i G_i(\boldsymbol{p}) + G_{n+1}(\boldsymbol{p}) = \sum_{i=1}^{n} \alpha_i \left( G_i(\boldsymbol{p}) + G_{n+1}(\boldsymbol{p}) \right),$$

$$V(\boldsymbol{\alpha}) \equiv \max_{\boldsymbol{p} \in \mathcal{P}} G(\boldsymbol{\alpha}, \boldsymbol{p}) = G(\boldsymbol{\alpha}, \boldsymbol{p}(\boldsymbol{\alpha})),$$

where $\boldsymbol{p}(\boldsymbol{\alpha})$ denotes the $\boldsymbol{p}$ that maximizes $G$ for a fixed $\boldsymbol{\alpha}$. Suppose $\boldsymbol{\alpha}^*$ is a solution to $V(\boldsymbol{\alpha}^*) = \min_{\boldsymbol{\alpha}} V(\boldsymbol{\alpha})$. Then, $\boldsymbol{p}(\boldsymbol{\alpha}^*)$ is the solution to the *max-min* problem in (8).

Due to Lemma 1, our optimization problem (5) can be solved in the following steps. The minimum operation in (6) is first replaced by the weighted sum of the $K$ common rates to yield:

$$\mathcal{L}(\boldsymbol{\mu}_0, \boldsymbol{p}) \equiv \left( \mu_0 \sum_{k=1}^{K} \frac{\mu_{0k}}{\mu_0} \sum_{i=1}^{M} R_{0k}^{(i)} + \sum_{k=1}^{K} \mu_k R_k \right) - \lambda \sum_{i=1}^{M} P^{(i)}$$

$$= \sum_{i=1}^{M} \left[ \mu_0 \left( \sum_{k=1}^{K} \frac{\mu_{0k}}{\mu_0} R_{0k}^{(i)} + \sum_{k=1}^{K} \frac{\mu_{\pi^{(i)}(k)}}{\mu_0} R_{\pi^{(i)}(k)}^{(i)} \right) - \lambda P^{(i)} \right], \quad (9)$$

with a nonnegative weight vector $\boldsymbol{\mu}_0 \equiv [\mu_{01}, \mu_{02}, \cdots, \mu_{0K}]$ that satisfies $\mu_{01} + \mu_{02} + \cdots + \mu_{0K} = \mu_0$. The second step is finding the optimal power policy $\boldsymbol{p}(\boldsymbol{\mu}_0)$ that maximizes $\mathcal{L}$ for a given $\boldsymbol{\mu}_0$. In the final step, we search for $\boldsymbol{\mu}_0^*$ that minimizes $\mathcal{L}(\boldsymbol{\mu}_0, \boldsymbol{p}(\boldsymbol{\mu}_0))$, and the corresponding $\boldsymbol{p}(\boldsymbol{\mu}_0^*)$ is the optimal power allocation that solves the original optimization problem (6). Because $\mathcal{L}(\mu_0, p(\mu_0))$ is continuous and convex for a feasible $\mu_0$, the global minimum in the last step can be efficiently found using convex optimization methods.

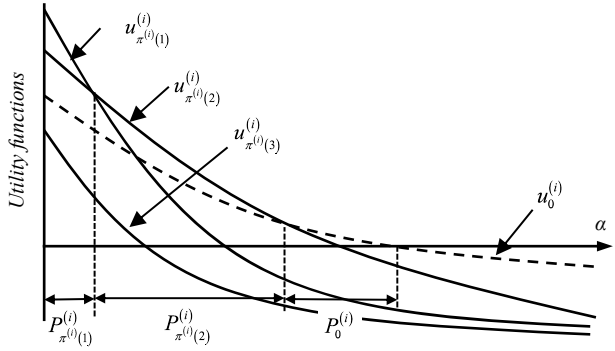

Fig. 2. An illustration of the utility functions $u_0^{(i)}(\alpha)$ and $u_{\pi^{(i)}(k)}^{(i)}(\alpha)$ for $k = 1, 2, 3$ (three users). In this example, the utility functions are properly ordered since $u_{\pi^{(i)}(1)}^{(i)}(\alpha)$ is maximum first, $u_{\pi^{(i)}(2)}^{(i)}(\alpha)$ is next, followed by $u_0^{(i)}(\alpha)$, until all utility functions are negative. The greedy power allocation strategy is also illustrated as $P_{\pi^{(i)}(1)}^{(i)}$, $P_{\pi^{(i)}(2)}^{(i)}$, and $P_0^{(i)}$. $P_{\pi^{(i)}(3)}^{(i)}$ is zero in this example.

For further simplification, we note that each summand in (9) for a fixed $i$ is maximized by $\boldsymbol{p}^{(i)} \equiv [P_0^{(i)}, P_{\pi^{(i)}(1)}^{(i)}, \ldots, P_{\pi^{(i)}(K)}^{(i)}]$ (power allocated to $i^{\text{th}}$ parallel channel) only. Hence, it suffices to consider the individual parallel channels separately in our optimization problem. The power split *across* $M$ parallel channels will be controlled by the Lagrange multiplier $\lambda$, which leads to a water-filling type algorithm.

With this observation, we now exclusively focus on the $i^{\text{th}}$ channel. Writing out the problem explicitly, we now have in the $i^{\text{th}}$ channel:

$$\max_{\boldsymbol{p}^{(i)}} \mu_0 \frac{N^{(i)}}{N} \left[ \sum_{k=1}^{K} \frac{\mu_{0k}}{\mu_0} \log_2 \left( 1 + \frac{P_0^{(i)}}{Z_k^{(i)} + \sum_{j=1}^{K} P_j^{(i)}} \right) \right.$$
$$\left. + \sum_{k=1}^{K} \frac{\mu_{\pi^{(i)}(k)}}{\mu_0} \log_2 \left( 1 + \frac{P_{\pi^{(i)}(k)}^{(i)}}{Z_{\pi^{(i)}(k)}^{(i)} + \sum_{j<k} P_{\pi^{(i)}(j)}^{(i)}} \right) \right] - \lambda P^{(i)}. \quad (10)$$

We use the utility functions method in [6] to solve the above optimization problem. Utility functions are defined as

$$u_{\pi^{(i)}(k)}^{(i)}(\alpha) \equiv \left( \frac{\mu_{\pi^{(i)}(k)}/\mu_0}{Z_{\pi^{(i)}(k)}^{(i)} + \alpha} \right) - \left( \frac{\lambda \ln 2}{\mu_0 N^{(i)}/N} \right); \quad k = 1, 2, \cdots, K,$$

$$u_0^{(i)}(\alpha) \equiv \left( \sum_{k=1}^{K} \frac{\mu_{0k}/\mu_0}{Z_k^{(i)} + \alpha} \right) - \left( \frac{\lambda \ln 2}{\mu_0 N^{(i)}/N} \right), \quad (11)$$

where $\ln(\cdot)$ denotes natural logarithm. These utility functions are all convex and monotonically decreasing in $\alpha$ with the same asymptote. By observing that $\ln(1 + P_2/(Z + P_1)) = \int_{P_1}^{P_1+P_2} 1/(Z+\alpha) d\alpha$, our optimization problem (10) may be expressed in terms of the utility functions as

$$\max_{\boldsymbol{p}^{(i)}} \frac{N^{(i)} \mu_0}{N \ln 2} \left[ \sum_{k=1}^{K} \int_{\sum_{j=1}^{k-1} P_{\pi^{(i)}(j)}^{(i)}}^{\sum_{j=1}^{k} P_{\pi^{(i)}(j)}^{(i)}} u_{\pi^{(i)}(k)}^{(i)}(\alpha) d\alpha + \int_{\sum_{j=1}^{K} P_j^{(i)}}^{P^{(i)}} u_0^{(i)}(\alpha) d\alpha \right] \quad (12)$$

and the optimal $\boldsymbol{p}^{(i)}$ is the one that maximizes the terms inside the square bracket. Fig. 2 illustrates the utility functions for three users. We let

$$u_{\max}^{(i)}(\alpha) \equiv \max \left\{ u_{\pi^{(i)}(1)}^{(i)}(\alpha), \cdots, u_{\pi^{(i)}(K)}^{(i)}(\alpha), u_0^{(i)}(\alpha), 0 \right\} \quad (13)$$

to denote an envelope of the utility functions and a zero function. Then, it is easy to see that (12) is upper-bounded by

$$\frac{N^{(i)} \mu_0}{N \ln 2} \int_0^{P^{(i)}} u_{\max}^{(i)}(\alpha) d\alpha. \quad (14)$$

This upper-bound can be achieved with an equality if the utility functions are *properly ordered*. That is, if we were to list the utility functions that attains the envelope function $u_{\max}^{(i)}(\alpha)$ over $\alpha > 0$, it must be in the order of $u_{\pi^{(i)}(1)}^{(i)}(\alpha)$, $u_{\pi^{(i)}(2)}^{(i)}(\alpha)$, …, $u_{\pi^{(i)}(K)}^{(i)}(\alpha)$, followed by $u_0^{(i)}(\alpha)$, although it is possible to skip any of the utility functions along the way. When properly ordered, the upper-bound is achieved by a greedy strategy by setting $P_{\pi^{(i)}(1)}^{(i)}$ equal to the length of the segment of $\alpha$ where $u_{\pi^{(i)}(1)}^{(i)}(\alpha)$ is maximum, setting $P_{\pi^{(i)}(2)}^{(i)}$ equal to the length where $u_{\pi^{(i)}(2)}^{(i)}(\alpha)$ is maximum, and so on, until all utility functions are negative or all power are exhausted. The utility functions as well as the greedy power allocation are illustrated in Fig. 2 for the 3-user example.

We now consider several cases where the utility functions are properly ordered and hence the upperbound (14) can be achieved via the greedy power allocation strategy. These cases exploit the following bounds on $u_0^{(i)}(\alpha)$:

$$\left( Z_{\pi^{(i)}(K)}^{(i)} + \alpha \right)^{-1} \leq \left( u_0^{(i)}(\alpha) + \frac{\lambda \ln 2}{\mu_0 N^{(i)}/N} \right) \leq \left( Z_{\pi^{(i)}(1)}^{(i)} + \alpha \right)^{-1},$$

due to $Z_{\pi^{(i)}(1)}^{(i)} \leq Z_{\pi^{(i)}(2)}^{(i)} \leq \cdots \leq Z_{\pi^{(i)}(K)}^{(i)}$.

*Case 1.* $\mu_0 < \min\{\mu_1, \mu_2, \cdots, \mu_K\}$:

In this case, since $u_0^{(i)}(\alpha) < u_{\pi^{(i)}(1)}^{(i)}(\alpha)$ for all $\alpha$, power is never allocated to the common message and sending only the independent messages is optimal in $i^{\text{th}}$ channel. The same holds for any $i$ and hence sending a common message in any of the channels incurs a capacity penalty. In general, more than 2-layers are needed for the independent messages.

*Case 2.* $\mu_0 \geq \max\{\mu_1, \mu_2, \cdots, \mu_K\}$:

In this case, $u_{\pi^{(i)}(K)}^{(i)}(\alpha) \leq u_0^{(i)}(\alpha)$ and hence the common message is always preferred over the independent message of the user with the highest noise power. Therefore, it is *possible* that the common message is transmitted, but whether it does or does not depends on the noise powers of the other users. In general, more than 2-layers are needed.

*Case 3.* $\mu_{\pi^{(i)}(1)} = \max\{\mu_1, \mu_2, \cdots, \mu_K\}$ and $\mu_0 \leq \mu_{\pi^{(i)}(1)}$:

In this case, since $u_0^{(i)}(\alpha) \leq u_{\pi^{(i)}(1)}^{(i)}(\alpha)$ for all $\alpha$, it is optimal to allocate all power to the independent messages. Furthermore, since $u_{\pi^{(i)}(1)}^{(i)}(\alpha)$ is larger than any of the other utility functions, transmitting to only the best user (the one with the lowest noise power) becomes optimal. That is, one-layer is optimal in $i^{\text{th}}$ channel.

*Case 4:* $\mu_{\pi^{(i)}(1)} = \max\{\mu_1, \mu_2, \cdots, \mu_K\}$ and $\mu_0 > \mu_{\pi^{(i)}(1)}$:

In this case, again $u_{\pi^{(i)}(1)}^{(i)}(\alpha)$ is larger than $u_{\pi^{(i)}(k)}^{(i)}(\alpha)$ for all $k \neq 1$ and the best user's message is selected over all other users' messages. This leaves only $u_0^{(i)}(\alpha)$ and $u_{\pi^{(i)}(1)}^{(i)}(\alpha)$, i.e., the common message and the independent message of the best user and 2-layers become optimal in $i^{\text{th}}$ channel.

As a special instance when $\mu_1 = \mu_2 = \cdots = \mu_K = \mu$ and $\mu_0 \leq \mu$, the conditions of Case 3 are satisfied for *all M* parallel channels and hence orthogonal multiplexing (of the independent messages of the best user in each of $M$ parallel channels) becomes optimal. On the other hand, when $\mu_1 = \mu_2 = \cdots = \mu_K = \mu$ and $\mu_0 > \mu$, the conditions of Case 4 are satisfied for all channels and hence two-layer superposition of the common message and the independent message of the best user (with the lowest effective noise power) in each of $M$ parallel channels becomes optimal. This completes the proof of Proposition 1. ∎

### C. Extension to Fading Channels with Partial CSI at the Transmitter

We now turn to fading GBC where the transmitter is assumed to know only the statistics of the channel, but not the individual channel realisations. Unfortunately, in the absence of CSI at the transmitter, the capacity region of fading GBC (with or without the common message) is not known in general. The difficulty stems from lack of stochastic orderings of the users due to non-degraded nature of the fading GBCs, which in turn breaks down our prior assumption that one of the users is able to decode and strip off another user's message. For this reason, the following natural extension to $\mathcal{C}$ in (3):

$$\mathcal{C}_f \equiv \bigcup_{p \in \mathcal{P}} \left\{ \boldsymbol{R} : R_0 \leq \min_{k \in [1:K]} \left[ \mathbb{E}\left(\sum_{i=1}^{M} R_{0k}^{(i)}\right) \right], \\ R_k \leq \sum_{i=1}^{M} \mathbb{E}\left(R_k^{(i)}\right); k \in [1:K] \right\}, \quad (15)$$

which is obtained by averaging the rates over the channel statistics cannot be considered achievable [37]. Achieving known inner-bounds (e.g., [38]) needs a considerably more sophisticated coding scheme than the superposition coding and in general involves joint encoding of more than two messages.

Fortunately, in a special case when individual fading distributions of the user channels are *of the same form*, $\mathcal{C}_f$ becomes the capacity region and the weighted sum-rate optimality of the two-layer superposition can be established as in the static channel case. More formally, suppose that $k^{\text{th}}$ user channel gain in the $i^{\text{th}}$ parallel channel can be written as

$$H_k^{(i)} = \sqrt{\sigma_{H_k^{(i)}}^2} \Phi_k^{(i)}, \quad (16)$$

where $\sigma_{H_k^{(i)}}^2$ is the variance of $H_k^{(i)}$ and $\Phi_1^{(i)}$, $\Phi_2^{(i)}$, ..., $\Phi_K^{(i)}$ are random variables with unit variances that have *same marginal probability distributions*. This includes the case when all users experience Rayleigh fading with the channel gains $H_k^{(i)} \sim \mathcal{CN}\left(0, \sigma_{H_k^{(i)}}^2\right)$. A long term average noise power may also be defined as $\bar{Z}_k^{(i)} \equiv \frac{N^{(i)}}{N} \sigma_k^2 / \sigma_{H_k^{(i)}}^2$. We now show that 2-layer superposition is weighted sum-rate optimal in this scenario.

*Proposition 2 (Optimality of two-layer superposition in a class of fading channels):* Suppose fading GBC has the individual fading channels satisfying (16) and only receivers have CSI and transmitter knows only the statistics of the channel. Then, in each of the $M$ parallel channels, sending the two-layered superposition of the common message and a single independent message of the user with the lowest long term average noise power can achieve the maximum weighted sum-rate $\mu_0 R_0 + \mu \sum_{k=1}^{K} R_k$ for any $\mu_0 > \mu$.

*Proof:* See appendix. ∎

## IV. Multicast-Unicast Multiplexing

One of the two main use-cases considered in this work is the Multicast-Unicast (MU) multiplexing where each user terminal with uplink capability receives both a common content (multicast) and an independent content (unicast). An important consideration in this use-case is to maximize the overall throughput of the network. Since the multicast content is received by all $K$ users, the network throughput is given by $K R_0 + \sum_{k=1}^{K} R_k$. Recognizing that this is a special case of the weighted sum-rate $\mu_0 R_0 + \mu \sum_{k=1}^{K} R_k$ with $\mu_0 > \mu$, the 2-layer superposition is optimal due to Proposition 1. A remaining problem is to find the optimal power allocation between the two layers and over $M$ parallel channels of the OFDM resources.

### A. Optimal Power Allocation for Two-Layer Superposition

Following Proposition 1, we exclusively consider the weighted sum-rate case: $\mu_1 = \cdots = \mu_K = \mu$ and $\mu_0 > \mu$, and find the optimal power allocation for the 2-layer superposition. Our goal here is to use the greedy power allocation strategy discussed in Section III.B, in order to achieve the upper-bound (14) to the weighted sum-rate. To this end, we first show that the two utility functions $u_{\pi^{(i)}(1)}^{(i)}(\alpha)$ and $u_0^{(i)}(\alpha)$ in (11) intersect at most once for $\alpha > 0$ and are properly ordered.

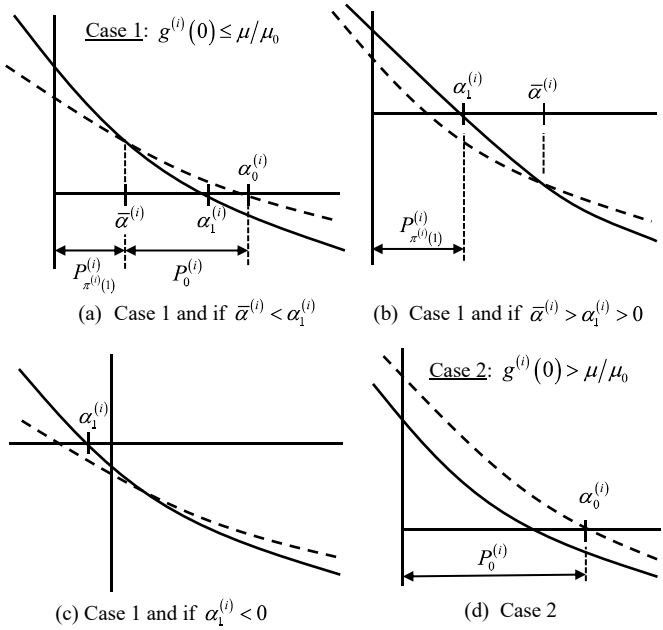

Fig. 3. Illustrations of the utility functions $u_0^{(i)}(\alpha)$ (in dashed lines) and $u_{\pi^{(i)}(1)}^{(i)}(\alpha)$ (in solid lines) in $\alpha$. Subfigures (a)-(c) correspond to Case 1 and (d) correspond to Case 2. Note that in Case 1, there is a single intersection between the utility functions in $\alpha > 0$. The optimal power allocations based on the greedy strategy are also illustrated in the figures as $P_0^{(i)}$ and $P_{\pi^{(i)}(1)}^{(i)}$.

First, let $\bar{\alpha}^{(i)} > 0$ be an intersection (if it exists) of the two utility functions, i.e., $u_{\pi^{(i)}(1)}^{(i)}(\bar{\alpha}^{(i)}) = u_0^{(i)}(\bar{\alpha}^{(i)})$ or

$$\sum_{k=1}^{K} \frac{\mu_{0k}/\mu_0}{Z_k^{(i)} + \bar{\alpha}^{(i)}} = \frac{\mu/\mu_0}{Z_{\pi^{(i)}(1)}^{(i)} + \bar{\alpha}^{(i)}}.$$

Multiplying both sides by $Z_{\pi^{(i)}(1)}^{(i)} + \bar{\alpha}^{(i)}$ (> 0) yields

$$\sum_{k=1}^{K} \frac{\mu_{0k}}{\mu_0} \frac{Z_{\pi^{(i)}(1)}^{(i)} + \bar{\alpha}^{(i)}}{Z_k^{(i)} + \bar{\alpha}^{(i)}} = \sum_{k=1}^{K} \frac{\mu_{0,\pi^{(i)}(k)}}{\mu_0} \frac{Z_{\pi^{(i)}(1)}^{(i)} + \bar{\alpha}^{(i)}}{Z_{\pi^{(i)}(k)}^{(i)} + \bar{\alpha}^{(i)}} = \frac{\mu}{\mu_0}. \quad (17)$$

Let $f(\alpha) \equiv (Z_1 + \alpha)/(Z_2 + \alpha)$ for $\alpha > 0$, where $Z_1$ and $Z_2$ are some nonnegative constants that satisfy $Z_2 > Z_1$. By noting that the first two derivatives of $f$ satisfy $f'(\alpha) > 0$ and $f''(\alpha) < 0$, $f$ is monotonically increasing in $\alpha$ and is strictly concave. Since a positive linear combination of strictly concave functions is also strictly concave, $g^{(i)}(\alpha)$ as defined by

$$g^{(i)}(\alpha) \equiv \sum_{k=1}^{K} \frac{\mu_{0,\pi^{(i)}(k)}}{\mu_0} \frac{Z_{\pi^{(i)}(1)}^{(i)} + \alpha}{Z_{\pi^{(i)}(k)}^{(i)} + \alpha} \quad (18)$$

is strictly concave in $\alpha > 0$ as well. Moreover, it is easy to see that $g^{(i)}$ is continuous and monotonically increasing in $\alpha$ towards its final value of 1. Therefore, $g^{(i)}$ can cross $\mu/\mu_0$ (<1) exactly once if it initially starts below $\mu/\mu_0$, i.e., if $g^{(i)}(0) \le \mu/\mu_0$, meaning that there is exactly one intersection between the two utility functions. Furthermore, in this case, $u_{\pi^{(i)}(1)}^{(i)}(\alpha)$ starts out smaller than $u_0^{(i)}(\alpha)$ and then becomes larger after the intersection, implying that the two utility functions are properly ordered. In the other case when $g^{(i)}(0) > \mu/\mu_0$, the utility functions do not intersect and $u_0^{(i)}(\alpha)$ is always larger than $u_{\pi^{(i)}(1)}^{(i)}(\alpha)$ for all $\alpha > 0$.

Since $g^{(i)}$ is monotonically increasing and differentiable, the intersection $\bar{\alpha}^{(i)}$ can be efficiently found by root-finding algorithms such as the Newton-Raphson's method. Moreover, the zeros of the utility functions, say $\alpha_0^{(i)}$ and $\alpha_1^{(i)}$ such that $u_0^{(i)}(\alpha_0^{(i)}) = 0$ and $u_{\pi^{(i)}(1)}^{(i)}(\alpha_1^{(i)}) = 0$, can be found similarly.

The structures of the optimal power allocation depend on whether $g^{(i)}(0) \le \mu/\mu_0$ or $g^{(i)}(0) > \mu/\mu_0$ and are summarized below and illustrated in Fig. 3.

*Case 1.* When $g^{(i)}(0) \le \mu/\mu_0$:

We obtain three subcases depending on the location of zeroes as illustrated in Fig. 3a-c. If $\bar{\alpha}^{(i)} < \alpha_1^{(i)}$ as in Fig. 3a, the two utility functions intersect before either function crosses zero. The greedy approach here is to allocate $P_{\pi^{(i)}(1)}^{(i)} = \bar{\alpha}^{(i)}$ and $P_0^{(i)} = \alpha_0^{(i)} - \bar{\alpha}^{(i)}$. If $\bar{\alpha}^{(i)} > \alpha_1^{(i)} > 0$ as in Fig. 3b, $u_{\pi^{(i)}(1)}^{(i)}$ crosses zero before the intersection and power is only allocated to the independent message, i.e., $P_{\pi^{(i)}(1)}^{(i)} = \alpha_1^{(i)}$ and $P_0^{(i)} = 0$. Lastly, if $\alpha_1^{(i)} < 0$ as in Fig. 3c, both utility functions are negative for all $\alpha > 0$ and no power is allocated in $i^{\text{th}}$ channel. The three different power allocations are expressed compactly as $P_{\pi^{(i)}(1)}^{(i)} = [\min\{\alpha_1^{(i)}, \bar{\alpha}^{(i)}\}]^+$ and $P_0^{(i)} = [\alpha_0^{(i)} - \bar{\alpha}^{(i)}]^+$.

*Case 2.* When $g^{(i)}(0) > \mu/\mu_0$:

In this case, there is no intersection between the two utility functions in $\alpha > 0$, and $u_0^{(i)}(\alpha) > u_{\pi^{(i)}(1)}^{(i)}(\alpha)$ for all $\alpha > 0$ as in Fig. 3d. Hence, it is optimal to allocate all power to the common message only, i.e., $P_0^{(i)} = [\alpha_0^{(i)}]^+$ and $P_{\pi^{(i)}(1)}^{(i)} = 0$.

The overall power allocation procedure is summarized in Algorithm 1. The algorithm first finds the greedy power allocation that maximizes the sum rate $\mathcal{L}(\mu_0, p)$ (9) for a particular weighting vector $\mu_0 = [\mu_{01}, \mu_{02}, \cdots, \mu_{0K}]$. Then, $\mu_0$ that minimizes $\mathcal{L}$ is found via a convex minimization. The corresponding $p$ is the globally optimal power allocation that solves the original optimization problem in (5) due to Lemma 1.

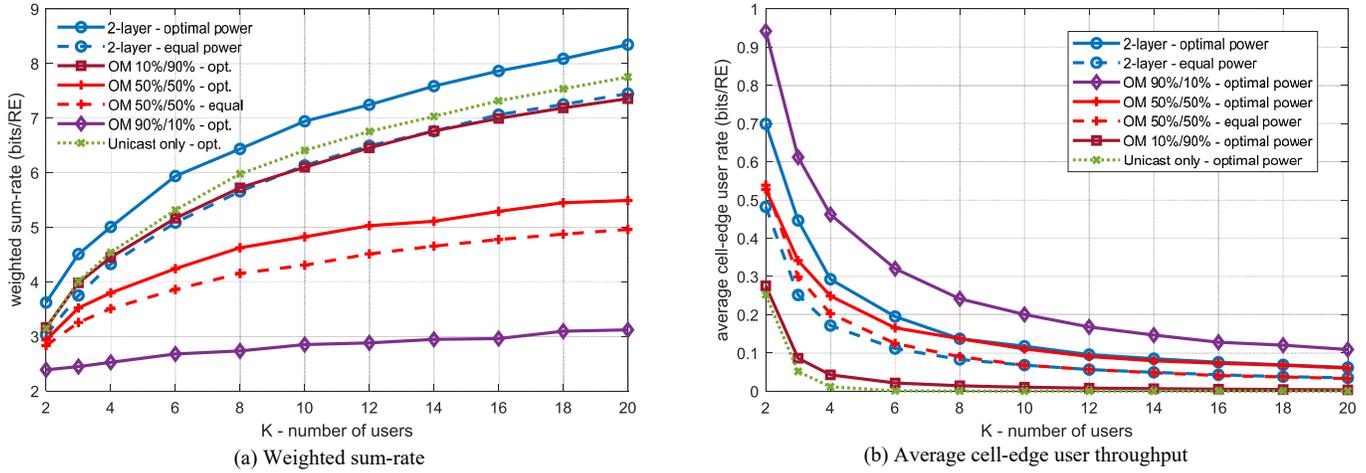

(a) Weighted sum-rate
(b) Average cell-edge user throughput

Fig. 4. Weighted sum-rates and average cell-edge user throughputs of the 2-layer superposition, the orthogonal multiplexing (OM) and the unicast-only transmission. Both optimal and equal power allocations are considered.

**Algorithm 1** Optimal Power Allocation over $M$ Parallel Channels of OFDMA in Multicast-Unicast Multiplexing

**STEP 1:** Initialize $\boldsymbol{\mu}_0 = [\mu_{01}, \mu_{02}, \cdots, \mu_{0K}]$ so that $\mu_{0k} \geq 0$ and $\sum_{k=1}^{K} \mu_{0k} = 1$.

**STEP 2 Gradual Power Water-filling and Greedy Power Allocation:**
Set $P_0^{(i)} = 0$ and $P_k^{(i)} = 0$ for $\forall i \in [1:M]$, $\forall k \in [1:K]$.
Set an initial waterline $1/\lambda$.

While $\left| \sum_{i=1}^{M} \left( P_0^{(i)} + \sum_{k=1}^{K} P_k^{(i)} \right) - P \right| < tolerance$,

  Adjust the waterline $1/\lambda$.
  For $i = 1, 2, \ldots, M$
    Compute $\bar{\alpha}^{(i)}$, $\alpha_0^{(i)}$, $\alpha_1^{(i)}$ such that $g^{(i)}(\bar{\alpha}^{(i)}) = \mu/\mu_0$, $u_0^{(i)}(\alpha_0^{(i)}) = 0$,
    and $u_{\pi^{(i)}(1)}^{(i)}(\alpha_1^{(i)}) = 0$ (using root-finding algorithms).

    If $g^{(i)}(0) \leq \mu/\mu_0$ then $P_{\pi^{(i)}(1)}^{(i)} = \left[ \min\{\alpha_1^{(i)}, \bar{\alpha}^{(i)}\} \right]^+$, $P_0^{(i)} = \left[ \alpha_0^{(i)} - \bar{\alpha}^{(i)} \right]^+$

    Else $P_{\pi^{(i)}(1)}^{(i)} = 0$, $P_0^{(i)} = \left[ \alpha_0^{(i)} \right]^+$

$\boldsymbol{p} = \left[ P_0^{(1)}, P_1^{(1)}, \ldots, P_K^{(1)}, P_0^{(2)}, P_1^{(2)}, \ldots, P_K^{(2)}, \ldots, P_0^{(M)}, P_1^{(M)}, \ldots, P_K^{(M)} \right]$

**STEP 3:** Compute the sum rate $\mathcal{L}(\boldsymbol{\mu}_0, \boldsymbol{p}) = \sum_{k=1}^{K} \mu_{0k} R_{0k}(\boldsymbol{p}) + \mu \sum_{k=1}^{K} R_k(\boldsymbol{p})$.

**STEP 4 Convex Minimization Iteration:**
Repeat STEPS 2-3 to search for $\boldsymbol{\mu}_0$ that minimizes $\mathcal{L}$ using a convex minimization algorithm.
Return $\boldsymbol{p}$ at the minimizer $\boldsymbol{\mu}_0$.

We note that the same algorithm may be used to obtain the optimal powers in fading channels due to Proposition 2, when the individual channel distributions satisfy the conditions of the proposition. In this case, the user orderings $\pi^{(i)}(1)$, $\pi^{(i)}(2)$, ..., $\pi^{(i)}(K)$ in the algorithm are based on the users' long-term average noise powers, i.e., in increasing order of $\bar{Z}_k^{(i)}$, and the sum rates are averaged over the fading distributions.

We note that in the case when all users have same channel qualities, Algorithm 1 allocates all available power to the common message. In another extreme when one of the users has relatively poor channel quality compared to other users, Algorithm 1 allocates little to no power to the common message and only the independent messages are sent.

### B. Numerical Results

We now present numerical evaluations of the weighted sum-rates of the MU multiplexing in static channels. We consider an OFDM composed of $M = 10$ parallel channels with equal numbers of REs. Average SNR of $k^{th}$ user in $i^{th}$ channel is defined as $P/\bar{Z}_k^{(i)}$. Users are uniformly placed in a circular cell with a realistic modelling of the signal propagation with a mean path loss model based on the Okumura-Hata suburban model, shadowing log-normal fading to model macroscopic fading as well as Rayleigh fading to model the multipath propagation in in-car (indoor) environment. The simulation assumptions for the transmitter (e.g. transmit power, effective antenna heights and diagrams), receiver (e.g. antenna height and gain, noise figure and bandwidth) and environment (e.g. indoor penetration loss) are extracted from Tables 1-3 in [46] with a radius of 7.5 km (representative of low-power low-tower LPLT networks in rural environments), 700 MHz carrier frequency, 10 MHz channel bandwidth and log-normal shadowing with 8 dB of standard deviation. All results are averaged over at least 5,000 channel realisations to observe a statistical mean behavior.

Fig. 4a shows the weighted sum-rate $KR_0 + \sum_{k=1}^{K} R_k$ when the 2-layer superposition and the OM are used. For the 2-layer, the optimal power allocation is obtained by Algorithm 1 and equal power allocation splits power equally among the $M$ channels and in between the unicast message of the best user and the common message component in each channel. For OM, we consider various channel assignments between multicast/unicast with optimal or equal power allocations. In order to exploit all degrees of freedom available to OM, we allow OM to dynamically assign channels depending on the channel states. For example, OM 10%/90% assigns 9 out of 10 best channels (i.e., with largest instantaneous channel realisations) to unicast first, then the remaining channel(s) to multicast, which yielded in our simulations strictly higher weighted sum-rates than static assignments of the channels or giving priority to multicast first. Also plotted in the figure is the optimal sum-rate of unicast-only transmission, which sends a single unicast message of the best user in each channel using the waterfilling algorithm for its power allocation [6]-[7].

We observe that the 2-layer superposition yields the largest weighted sum-rate and the rate gain increases with the number of users, demonstrating the multicast gain. We note that using additional layers cannot improve this sum-rate due to Proposition 1. Comparing with the unicast-only case, the benefit to incorporating the multicast-unicast multiplexing is apparent. Interestingly, the 2-layer with the equal power allocation is also seen to yield noticeable gains over OM, which may be attributed to the multiuser diversity in OFDMA systems where only subchannels with good channel gains are assigned to each user (cf. [39]). The good performance of the equal power allocation implies that the transmitter can realize significant gains by knowing only the user orderings (and not the precise channel gains).

The sum-rate shows only an aggregated network performance. In practical cellular systems, performance of cell-edge users is an important performance metric related to fairness and coverage area. Fig. 4b plots the average cell-edge throughput ($R_0$ plus unicast rate to the cell-edge user) vs. $K$. The "cell-edge user" here refers to the user with the lowest instantaneous channel realisation[6]. The cell-edge rates decrease with increasing $K$ since the chance that one of the users suffers from a deep fade increases with $K$ and the total available power must be shared between user messages.

We observe that OM 90%/10% yields higher rate to the cell-edge user than that of the 2-layer superposition. In fact, the cell-edge rate would be maximized by allocating all resources to multicast. However, this small gain in the cell-edge rate comes at a rather large penalty in the rates of all other users – as evidenced by the small weighted sum-rate of OM 90%/10% in Fig. 4a. On the other hand, the 2-layer provides considerable rate to the cell-edge user (higher than OM 50%/50% and 10%/90%), while providing the highest possible overall sum-rate, hence yielding more efficient use of spectrum than OM. We also note that the cell-edge rate of the unicast-only case is always near zero since the best user is always preferred over the cell-edge user if the sum-rate is to be maximized.

## V. Broadcast-Unicast Multiplexing

The Broadcast-Unicast (BU) Multiplexing models a base station that transmits a common message to an unknown number of receive-only terminals (without uplink capabilities) and additionally transmits independent messages to $K$ users with uplink capabilities (say, unicast users) that also receive the common content. The primary goal here is to serve a certain percentage (say 98%) of the receive-only terminals in the cell, while opportunistically providing unicast messages to the unicast users. We show below that two-layer superposition is sum-rate optimal in the BU use-case.

---

[6] Practical deployments use the $x$% percentile (where $x$=5 is a common assumption) of the user spectral efficiency for the cell-edge user (i.e. the $x$% of users with the worst spectral efficiencies are discarded) instead of the worst user in all the area, which would provide higher spectral efficiencies than those reported in Fig. 4b.

### A. Sum-Rate Optimality of Two-Layer Superposition and the Modified Optimal Power Allocation Algorithm

First, the objective function (6) is updated to incorporate a new rate constraint $R_0 \geq r_0$, where $r_0$ is set to the minimum link capacity of a certain percentage of receive-only devices ($r_0$ is typically determined from field tests prior to network deployment). The new objective function is:

$$\max_{\substack{p \in \mathcal{P} \\ R_0 \geq r_0}} \left\{ \mu_0 \min_{k \in [1:K]} \left\{ \sum_{i=1}^{M} R_{0k}^{(i)} \right\} + \sum_{k=1}^{K} \mu_k R_k - \lambda \sum_{i=1}^{M} P^{(i)} + \lambda_0 \min_{k \in [1:K]} \left\{ \sum_{i=1}^{M} R_{0k}^{(i)} \right\} \right\}, \quad (19)$$

where both $\lambda$ and $\lambda_0$ are nonnegative Lagrange multipliers. Setting $\bar{\mu}_0 \equiv \mu_0 + \lambda_0$ as a new rate reward for the common message, (19) can be simplified as

$$\max_{\substack{p \in \mathcal{P} \\ R_0 \geq r_0}} \left\{ \bar{\mu}_0 \min_{k \in [1:K]} \left\{ \sum_{i=1}^{M} R_{0k}^{(i)} \right\} + \sum_{k=1}^{K} \mu_k R_k - \lambda \sum_{i=1}^{M} P^{(i)} \right\}, \quad (20)$$

which is same as our original objective function (6) with the new rate reward $\bar{\mu}_0$. This indicates that the same Algorithm 1 in Section IV.A finds the optimal power allocation, but with increased rate reward to the broadcast message until the new constraint $R_0 \geq r_0$ is satisfied. Also, since $\bar{\mu}_0 \geq \mu_0$, Proposition 1 applies, i.e., the 2-layer superposition remains optimal.

### B. Numerical Results

The simulation setup and assumptions for the BU use-case are the same as those of the MU use-case in Section IV.B, but in a car-mounted environment [46] to model outdoor receivers with log-normal shadowing fading with 5.5 dB of standard deviation. The broadcast network is designed to deliver a target broadcast rates of 2 bits/RE that corresponds to a coverage area of 99.7% for the considered 7.5 km cell radius. We consider the fading channel model, where the channel gains are Rayleigh distributed and change independently over each symbol. To represent receive-only terminals, one additional "cell-edge" user with the link capacity $C_B$ = 2 bits/RE (corresponding to the 99.7% percentile of the user spectral efficiency) is inserted into the network of $K_u$ unicast users. Allowing this additional user to recover the common message allows all receive-only terminals to recover the same message, assuming that the channel is stochastically degraded. A new rate constraint $R_0 \geq r_0$ is imposed where $r_0$ is expressed as a fraction of $C_B$. The optimal power allocation with the rate constraint is obtained by Algorithm 1 with the new rate reward $\bar{\mu}_0$. We assume that the transmitter only knows the average SNR of the unicast users and not their instantaneous channel realisations. As before, we consider an OFDM with $M$=10 parallel channels with same number of REs per channel.

Fig. 5a shows the maximum sum-rate $R_0 + \sum_{k=1}^{K_u} R_k$ of the 2-layer superposition when $R_0$ is constrained to be at least 90%, 70%, or 50% of $C_B$. We observe that the 2-layer

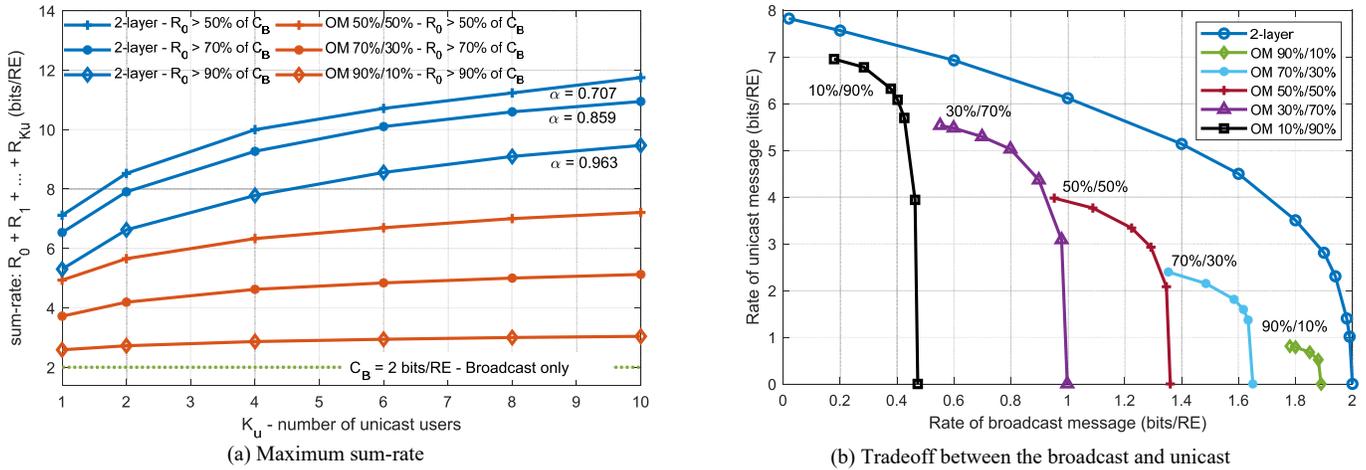

(a) Maximum sum-rate  (b) Tradeoff between the broadcast and unicast

Fig. 5. Performance in Broadcast-Unicast Multiplexing use-case. (a) Maximum sum-rates of the 2-layer superposition and the OM, with common message rates constrained to be at least 90%, 70% or 50% of broadcast-only capacity. (b) Tradeoff between the broadcast and unicast for 2-layer superposition and OM.

superposition is able to provide significant unicast rates while meeting the minimum broadcast rate constraints for the receive-only terminals. Also plotted is the maximum sum-rate of OM with various channel assignment between the broadcast and the unicast messages. By comparing with the 2-layer case, we observe that at the same $R_0$, more unicast message rates can be supported by the 2-layer than the OM. Also for a small tradeoff in $R_0$, significant overall capacity gain can be obtained. For instance, the sum-rate increases from 2.0 to 9.5 at $K_u = 10$ when 10% of the broadcast rate is traded off.

The tradeoff between the broadcast and unicast message rates is shown more explicitly in Fig. 5b, which shows the maximum unicast rate of a single unicast user with varying broadcast message rates. Both two-layer and OM with various channel assignments are considered. In addition to providing overall higher rates than the OM, the 2-layer also offers greater flexibility in tradeoffs between the broadcast and the unicast. Moreover, the steep slope near the x-axis implies that for a small tradeoff in the broadcast rate, a large unicast message rate can be supported.

## VI. TRANSCEIVER DESIGN BASED ON 3GPP LTE-ADVANCED PRO SPECIFICATIONS AND NUMERICAL EVALUATIONS

In this section, we present a transceiver design based on the 3GPP technical specifications of LTE-A Pro physical layer. We focus on LTE eMBMS [40], which can multiplex unicast and broadcast signals in the same carrier by (orthogonal) time division multiplexing (TDM). Unicast and broadcast signals are conveyed in separate transport channels called Physical Downlink Shared Channel (PDSCH) and Physical Multicast Channel (PMCH), respectively, which are allocated to different sub-frames (with a duration of 1ms). While the sub-frames allocated to broadcast (PMCH) use the full signal bandwidth, the sub-frames used for unicast (PDSCH) can orthogonally allocate various unicast users according to specific radio resource management (RRM) procedures. We consider the TDM of broadcast/multicast and unicast signals in LTE-A Pro as the baseline performance that will be compared with the performance provided by our proposed two-layered broadcast/multicast and unicast superposition transmission (hereafter called BMUST) system, which is detailed next.

### A. Two-Layer Broadcast/Multicast and Unicast Superposition Transmission (BMUST) Architecture with LTE-A Pro

The BMUST transceiver architecture is shown in Fig. 6. The transmitter shown in Fig. 6a takes $K+1$ information blocks from upper layers, i.e. Transport Blocks (TBs), as specified in TS 36.213 (Physical Layer procedures) [41] to be conveyed by the PMCH and PDSCH transport channels. The Modulation and Coding Scheme (MCS) settings selected are based on table 7.1.7.1-1A, which include QPSK, 16QAM, 64QAM and 256QAM constellations. The TBs are processed by the multiplexing and channel coding procedures (as specified in TS 36.212 [42]) where the output coded bits are scrambled and modulated to normalized complex-valued symbols from the available modulation schemes (as specified in TS36.211 (Physical Channels and Modulation) [35]). The PDSCH transport channels from the $K$ unicast users are mapped to REs according to a specific RRM algorithm. In this evaluation the unicast signals are equally allocated to the resources in the sub-frame. For BMUST the PMCH and $K$ PDSCH transport channels are combined with a scaling factor $\sqrt{\alpha}$ and $\sqrt{1-\alpha}$, respectively, where $\alpha \in [0,1]$ to maintain the transmit power constraint (i.e. lower $\alpha$ factors allocate more power to the unicast signals and high $\alpha$ factors allocate more power to broadcast/multicast signals). Finally, the superimposed symbols are mapped to the REs in the time-frequency grid and reference signals (RS) are introduced as per TS36.211 before OFDM modulation for transmission on a single transmit antenna configuration.

For fair comparison between all the considered configurations, we use the same RS pattern as defined for antenna port 4. We also note that for BMUST there is no superposition on the REs containing RS. The BMUST receiver shown in Fig. 6b at the $j$th user, after performing OFDM demodulation from the received signal and estimating the

This is the author's version of an article that has been published in this journal. Changes were made to this version by the publisher prior to publication. The final version of record is available at http://dx.doi.org/10.1109/TWC.2019.2950205Copyright (c) 2019 IEEE. Personal use is permitted. For any other purposes, permission must be obtained from the IEEE by emailing pubs-permissions@ieee.org.

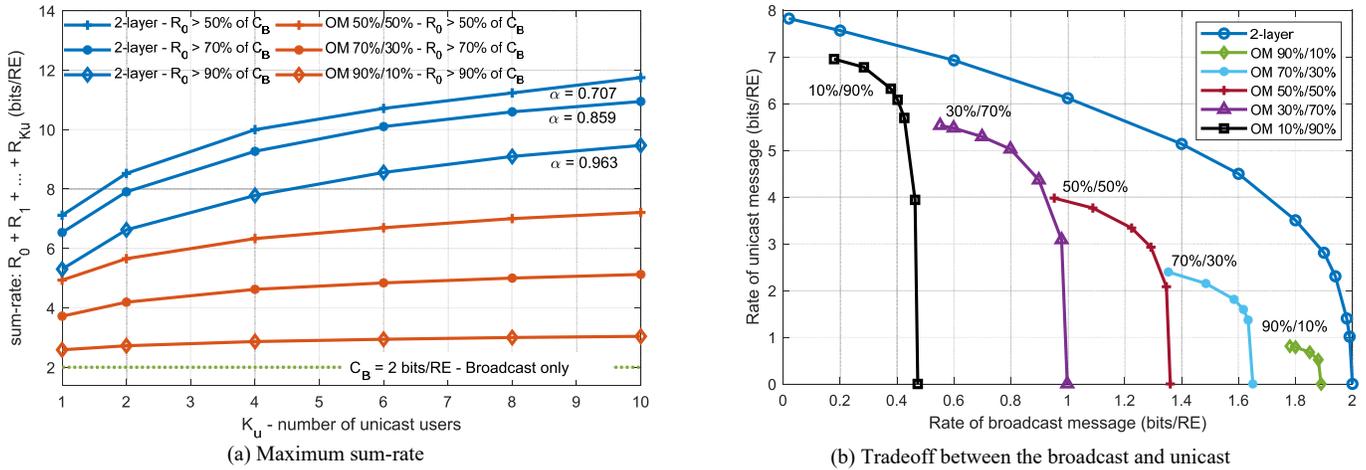

(a) Maximum sum-rate      (b) Tradeoff between the broadcast and unicast

Fig. 5. Performance in Broadcast-Unicast Multiplexing use-case. (a) Maximum sum-rates of the 2-layer superposition and the OM, with common message rates constrained to be at least 90%, 70% or 50% of broadcast-only capacity. (b) Tradeoff between the broadcast and unicast for 2-layer superposition and OM.

superposition is able to provide significant unicast rates while meeting the minimum broadcast rate constraints for the receive-only terminals. Also plotted is the maximum sum-rate of OM with various channel assignment between the broadcast and the unicast messages. By comparing with the 2-layer case, we observe that at the same $R_0$, more unicast message rates can be supported by the 2-layer than the OM. Also for a small tradeoff in $R_0$, significant overall capacity gain can be obtained. For instance, the sum-rate increases from 2.0 to 9.5 at $K_u = 10$ when 10% of the broadcast rate is traded off.

The tradeoff between the broadcast and unicast message rates is shown more explicitly in Fig. 5b, which shows the maximum unicast rate of a single unicast user with varying broadcast message rates. Both two-layer and OM with various channel assignments are considered. In addition to providing overall higher rates than the OM, the 2-layer also offers greater flexibility in tradeoffs between the broadcast and the unicast. Moreover, the steep slope near the x-axis implies that for a small tradeoff in the broadcast rate, a large unicast message rate can be supported.

## VI. TRANSCEIVER DESIGN BASED ON 3GPP LTE-ADVANCED PRO SPECIFICATIONS AND NUMERICAL EVALUATIONS

In this section, we present a transceiver design based on the 3GPP technical specifications of LTE-A Pro physical layer. We focus on LTE eMBMS [40], which can multiplex unicast and broadcast signals in the same carrier by (orthogonal) time division multiplexing (TDM). Unicast and broadcast signals are conveyed in separate transport channels called Physical Downlink Shared Channel (PDSCH) and Physical Multicast Channel (PMCH), respectively, which are allocated to different sub-frames (with a duration of 1ms). While the sub-frames allocated to broadcast (PMCH) use the full signal bandwidth, the sub-frames used for unicast (PDSCH) can orthogonally allocate various unicast users according to specific radio resource management (RRM) procedures. We consider the TDM of broadcast/multicast and unicast signals in LTE-A Pro as the baseline performance that will be compared with the performance provided by our proposed two-layered broadcast/multicast and unicast superposition transmission (hereafter called BMUST) system, which is detailed next.

### A. Two-Layer Broadcast/Multicast and Unicast Superposition Transmission (BMUST) Architecture with LTE-A Pro

The BMUST transceiver architecture is shown in Fig. 6. The transmitter shown in Fig. 6a takes $K+1$ information blocks from upper layers, i.e. Transport Blocks (TBs), as specified in TS 36.213 (Physical Layer procedures) [41] to be conveyed by the PMCH and PDSCH transport channels. The Modulation and Coding Scheme (MCS) settings selected are based on table 7.1.7.1-1A, which include QPSK, 16QAM, 64QAM and 256QAM constellations. The TBs are processed by the multiplexing and channel coding procedures (as specified in TS 36.212 [42]) where the output coded bits are scrambled and modulated to normalized complex-valued symbols from the available modulation schemes (as specified in TS36.211 (Physical Channels and Modulation) [35]). The PDSCH transport channels from the $K$ unicast users are mapped to REs according to a specific RRM algorithm. In this evaluation the unicast signals are equally allocated to the resources in the sub-frame. For BMUST the PMCH and $K$ PDSCH transport channels are combined with a scaling factor $\sqrt{\alpha}$ and $\sqrt{1-\alpha}$, respectively, where $\alpha \in [0,1]$ to maintain the transmit power constraint (i.e. lower $\alpha$ factors allocate more power to the unicast signals and high $\alpha$ factors allocate more power to broadcast/multicast signals). Finally, the superimposed symbols are mapped to the REs in the time-frequency grid and reference signals (RS) are introduced as per TS36.211 before OFDM modulation for transmission on a single transmit antenna configuration.

For fair comparison between all the considered configurations, we use the same RS pattern as defined for antenna port 4. We also note that for BMUST there is no superposition on the REs containing RS. The BMUST receiver shown in Fig. 6b at the $j$th user, after performing OFDM demodulation from the received signal and estimating the channel $H$ with the received RS, performs three processing stages to access the information in the PMCH and PDSCH transport channels. First, the receiver applies de-mapping, de-scrambling followed by de-multiplexing and channel decoding to access the information in the PMCH. Once the PMCH has been successfully decoded, it needs to be re-encoded, scaled according to the channel realisation and normalized by the factor $\sqrt{\alpha}$ before it can be subtracted from the main received signal. At this point, the PDSCH can be de-mapped, de-scrambled, de-multiplexed and channel decoded with reduced interference of the PMCH transport channel.

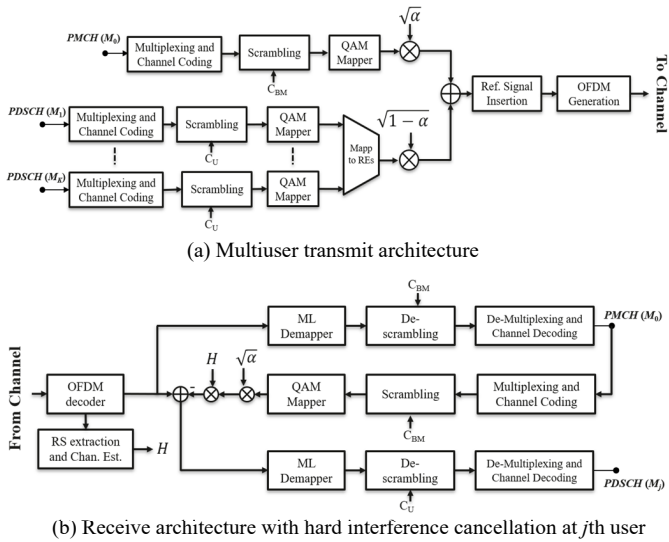

Fig. 6. Two-Layer superposition of Broadcast/Multicast (PMCH) and Unicast (PDSCH) signals with LTE-Advanced Pro physical layer.

The OM scheme can be described as a particularization of the BMUST architecture. At the transmitter, each sub-frame transmits either the PMCH or the $K$ PDSCH transport channels (according to the configured allocation ratio) with $\alpha = 1$ or $\alpha = 0$, respectively. The OM receiver can decode at each sub-frame the corresponding transport channel without any interference.

*B. System Parameters, Channel Model, and Simulation Assumptions for BMUST and OM Evaluation*

The selected bandwidth in the evaluations for the transmit signal is 10 MHz (50 Resource Blocks - RBs) and both transport channels PDSCH and PMCH use a single transmit antenna that correspond to transmission modes 1 (Downlink Control Format indicator 1) and 9 (Downlink Control Format indicator 1A), respectively. Both transport channels use a subcarrier spacing of 15 kHz with extended cyclic prefix duration of 16.7 µs with the addition of two OFDM symbols used for control in each sub-frame. As in Section V.B, it is assumed that the transmitter only knows the average SNR of the unicast users (e.g. CQI reporting). Hence, for both OM and BMUST, the transmitter does not schedule unicast signals across OFDMA subcarriers, i.e. unicast signals are allocated to the same RBs. We evaluate the performance of the considered schemes in a realistic mobile environment with a maximum Doppler spread of 220 Hz with multipath propagation. We use in the evaluations the Tapped Delay Line (TDL) 5G channel model (configuration B) defined by 3GPP in TR38.901 [43] with non-line of sight (NLOS) between the transmitter and receiver where the Doppler spread of each tap is characterized by the classical Jakes spectrum shape. The root mean square Delay Spread (DS) of the TDL models can be configured to characterize different propagation environments and in this paper we use the largest DS of 1µs defined in [43] that is categorized as "very long delay spread" and derived from measurements. The performance results in this section are shown against the average SNR as defined in Section IV.B. The receiver uses two receive antennas with a maximum ratio combiner (MRC) [cf. footnote 3], a maximum likelihood (ML) de-mapper and the algorithm used for the turbo decoder is a max-log-MAP decoding with a maximum of eight iterations. The receiver performs realistic estimation of the channel based on the received RS with independent one-dimensional linear interpolator in frequency domain followed by linear interpolation in time domain. Results with the assumption of ideal channel estimation are also provided. We note that while the performance with the assumption of ideal channel estimation provides optimistic results, the results with real channel estimation using a very simple linear interpolation provide pessimistic results. It is expected that a commercial receiver implementing more sophisticated channel estimation architectures would provide an estimation performance that lies between the two types of channel estimation assumptions included in this paper. The BMUST receiver performs ML detection of PMCH treating the faded unicast interference as (scaled) Gaussian noise.

*C. Broadcast/Multicast Numerical Evaluation with LTE-A Pro PMCH Transport Channel*

Fig. 7a presents the broadcast/multicast (PMCH) performance evaluation in terms of transport block error rate (BLER) vs. SNR (dB) for OM and BMUST for two different operation points with target spectral efficiencies of approximately 1.0 and 2.1 bits/RE with the assumption of ideal channel estimation. The target spectral efficiency of 1.0 bits/RE for the broadcast layer is also considered in Fig. 5a in Section V.B (i.e., with the configurations $R_0 > 50\%$ of $C_B$). The results show that BMUST provides a gradual performance degradation from the reference case ($\alpha = 1.0$) to the performance provided by OM for similar SE. However, robust MCS provides more flexibility to the BMUST system to trade off bit-rates (or coverage area) between the two layers carrying broadcast/multicast signals and unicast signals.

In particular for Fig. 7a, OM uses for PMCH all the frequency resources and half of the radio sub-frames (the other half of sub-frames are reserved for unicast) with MCS 6 & 14 while BMUST uses all time and frequency resources with MCS 3 & 7 and different powers allocated to PMCH (indicated in the figure with the factor $\alpha$). Given this resource allocation, the average SE in bits/RE provided by BMUST is 1.01 and 2.26 (for MCS 3 and 7, respectively), and for OM is 0.98 and 2.11 (for MCS 6 and 14, respectively), therefore the results are comparable. For reference, the curves with $\alpha = 1.0$ show the

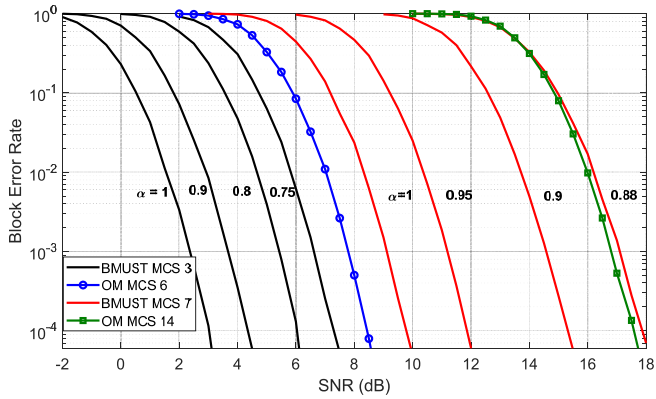

(a) Transport Block Error Rate vs. SNR (dB) with ideal channel estimation.

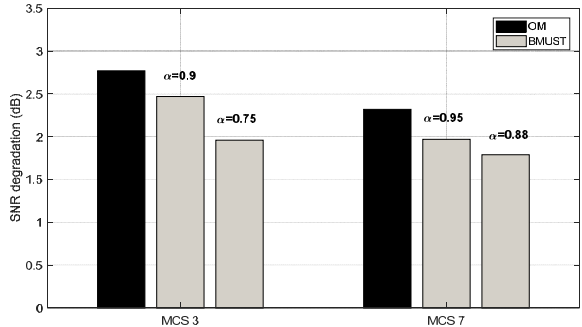

(b) SNR performance degradation (dB) due to real channel estimation compared to ideal estimation for OM and BMUST at target BLER=$10^{-3}$.

Fig. 7. Broadcast/Multicast (PMCH) performance for OM and non-orthogonal multiplexing (BMUST) at target spectral efficiencies of 1.0 and 2.1 bits/RE.

performance of the system with all the power, time and frequency resources allocated to PMCH (i.e. no unicast transmissions). The $\alpha$ factors are chosen to provide a performance that lays between the performance of $\alpha = 1$ and OM.

The rate of performance degradation with decreasing $\alpha$ is different for the two studied spectral efficiencies. For instance, the loss with $\alpha = 0.9$ at MCS 3 and 7 are approximately 1.3 dB and 5.5 dB, respectively. In this evaluation, the receiver decodes the broadcast/multicast signal assuming that the faded unicast interference is a Gaussian noise-like signal (although it is not) and hence with this type of receiving strategy MCS 3 can tolerate higher levels of interference than MCS 7.

Fig. 7b shows the PMCH performance degradation in terms of SNR (dB) at the target BLER of $10^{-3}$ due to the channel estimation error for MCS 3 and MCS 7. Overall, the results show similar performance degradation for both OM and BMUST. The results demonstrate that the imperfect estimation of the (self) interference power from the second layer carrying unicast signals does not have a noticeable impact on the performance of the first layer carrying broadcast/multicast signals. Particularly, in this study we compare OM and BMUST with the same MCS to analyze the robustness of the OM and BMUST against channel estimation errors with the same modulation and coding structure. With the considered real channel estimation algorithm, the estimation error reduces with increasing SNR since it reduces the noise at the received RS. This explains why for a given MCS (either 3 or 7) BMUST with $\alpha < 1$, with higher SNR operation point at BLER $10^{-3}$, has lower degradation due to realistic channel estimation.

### D. Unicast Numerical Evaluation with LTE-A Pro PDSCH Transport Channel

Fig. 8a shows the unicast (PDSCH) performance evaluation in terms of average SE (bits per RE) vs. the minimum SNR (dB) required to achieve a BLER equal to $10^{-3}$ for OM and BMUST with the assumption of ideal knowledge of the channel. Overall the results show that BMUST can provide significant gains in terms of increased SE or enhanced coverage area for users with high SNR where the gain increases with the user SNR.

In Fig. 8a each user is assigned five RBs (0.9 MHz) in the sub-frames allocated to unicast, which would allow the scheduling of 10 users in each sub-frame. One of the 10 unicast users is decoded and the performance is shown in Fig. 8a. For each curve, the following MCS are evaluated: 4, 6, 8, 10, 12, 14, 16, 17, 20, 22 and 24. Fig. 8a shows two OM configurations with 50% and 30% of the sub-frames allocated to unicast while the rest of the sub-frames reserved for the transmission of broadcast/multicast. For BMUST, all the sub-frames allocate unicast signals but with two values of power allocation factor $\alpha = \{0.75, 0.9\}$ and the broadcast/multicast layer uses MCS 3. For reference, the performance when no power (or sub-frame) is allocated to broadcast/multicast is also included (BMUST $\alpha = 0$ or equivalently OM unicast 100%). Comparing BMUST with $\alpha = 0.75$ against OM with 50% of sub-frames allocated to unicast, in terms of SE gains, BMUST can provide an improvement around 80% (at 31 dB of SNR) and in terms of SNR gains, BMUST can provide an SNR gain of 5.2 dB (at an SE of 3.5 bits/RE). If we also consider the performance of the broadcast/multicast layer in Fig. 7a, we can see that BMUST provides and SNR gain of 1.1 dB compared to OM. The superior performance of BMUST $\alpha = 0.75$ over OM unicast 50% is consistent with the theoretical results in Fig. 5a with an optimal $\alpha$ factor of 0.707 (Algorithm 1) and the same SE target for the broadcast layer. If the network requires better performance for the broadcast/multicast layer this could be achieved by trading-off some of the unicast gain by increasing the power allocation factor $\alpha$.

Fig. 8b studies the impact of real channel estimation in the performance of the unicast (PDSCH) signals for OM (with 50% of sub-frame allocation to unicast) and BMUST (with $\alpha = 0.75$). Due to channel estimation errors, the cancellation of the broadcast/multicast (PMCH) signal prior to decoding the unicast signals (PDSCH) is not perfect and this figure investigates the impact on the performance of BMUST. The

Table I. SNR degradation (dB) due to real vs. ideal channel estimation for OM with 50% sub-frames allocated to unicast and BMUST with $\alpha = 0.75$.

| OM: MCS, (spectral eff.), [ideal SNR (dB) at BLER $10^{-3}$] | | | | | | | |
|---|---|---|---|---|---|---|---|
| 4, (0.7), [5.8] | 8, (1.2), [10.4] | 12, (1.8), [14.6] | 16, (2.4), [19.2] | 17, (2.6), [21.2] | 20, (3.1), [23.1] | 22, (3.4), [25.8] | 24, (3.8), [31.3] |
| 2.3 | 1.9 | 1.8 | 2.0 | 2.2 | 2.6 | 3.7 | > 10 |
| BMUST: MCS, (spectral eff.), [ideal SNR (dB) at BLER $10^{-3}$] | | | | | | | |
| 4, (1.4), [11.8] | 8, (2.5), [16.3] | 12, (3.6), [20.5] | 16, (4.9), [25.3] | 17, (5.3), [27.3] | 20, (6.1), [29.0] | 22, (6.9), [31.8] | 24, (7.6), [37.3] |
| 2.2 | 2.0 | 2.2 | 3.5 | 5.0 | > 10 | > 10 | > 10 |

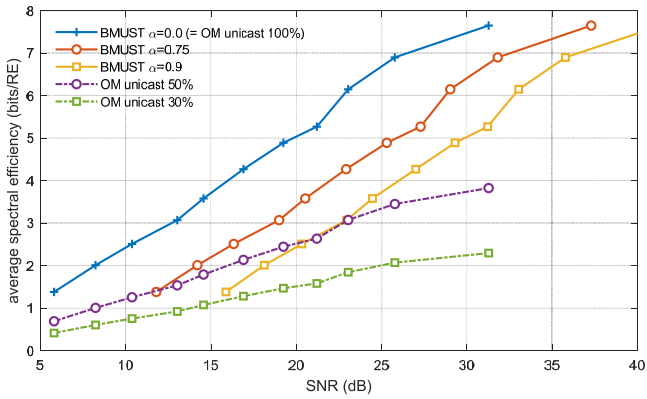

(a) Performance with ideal channel estimation. OM with 50% or 30% of the sub-frames allocated to unicast and BMUST using all sub-frames to transmit unicast with different $\alpha$.

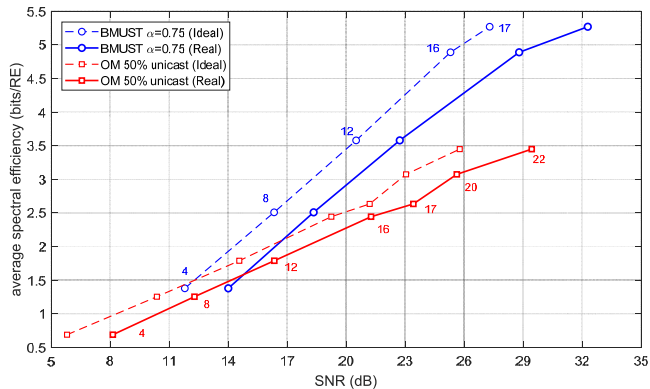

(b) Performance with real and ideal channel estimation. OM with 50% sub-frames allocated to unicast and BMUST using all sub-frames to transmit unicast with $\alpha = 0.75$.

Fig. 8. Unicast (PDSCH) performance for OM and BMUST. Average spectral efficiency (bits/RE) vs. the SNR (dB) required to achieve BLER $10^{-3}$.

results show similar degradation in the performance due to real channel estimation for both OM and BMUST where the effect of the imperfect cancellation due to real channel estimation has not shown significant performance impact. The results with a realistic (and simple receiver) still report significant gains in terms of SE and significant SNR gains.

Particularly, Fig. 8b includes only a subset of MCS of Fig. 8a (as indicated in the figure). Detailed degradation values for the simulated MCS are provided in Table I. We note that for both OM and BMUST the SNR degradation increases especially for MCS with high SNR operating point, e.g. BMUST with MCS 16 and OM with MCS 22 with an (ideal) SNR operating point around 26 dB both have a degradation of 3.5 dB in SNR. This is because for the evaluated environment with the considered channel estimation algorithm, the estimation error is near the maximum tolerable noise level of the transmission scheme. MCS configurations with (ideal) SNR operating points beyond 29 dB have a degradation larger than 10 dB for both OM and BMUST. For lower MCS, with lower SNR operating points, the estimation error is well below the tolerable noise level and the schemes have similar SNR degradations in the order of 2 dB. Regarding the gains provided by BMUST with real channel estimation, in terms of SE gains BMUST can provide an improvement around 41% at 29 dB of SNR, and in terms of SNR gains BMUST can provides even higher SNRs gain of 6.7 dB at an SE of 3.5 bits/RE (due to the higher degradation of OM at high SNR operation point). We note that we have considered a channel model with a very long delay spread and a receiver architecture with very simple channel estimation algorithm. Common scenarios with shorter delay spreads and/or receivers with better channel estimation algorithms will reduce the channel estimation error allowing the use of MCS with higher SNR operating points providing higher gains (closer to those reported in Fig. 8) for BMUST in terms in SE.

## VII. CONCLUSION

In this paper we demonstrate that non-orthogonal multiplexing of broadcast/multicast (targeting the cell-edge user) and unicast signals provides significant performance improvements over conventional orthogonal multiplexing from both the theoretical and practical points of view. In fact, we demonstrate that the considered two-layered superposition of broadcast/multicast and unicast signals in the downlink multiuser OFDMA systems is sum-rate optimal when the common message has higher priority than the individual unicast messages, and develop optimal power allocation across the two-layers and OFDMA resources. Given the recent interest by the mobile industry to enhance the broadcast/multicast capability in 3GPP standards (e.g. LTE eMBMS enhancements in Release-14 and Release-16 to enable the provision of TV services [44]) and the introduction of NOM schemes into the technical specifications (e.g. MUST in 3GPP for two unicast messages), our two-layer transceiver design for broadcast/multicast and unicast superposition based on the LTE-A-Pro physical layer (BMUST) shows significant performance improvements over orthogonal architectures specified in 3GPP while reusing existing interference cancellation capability available in the receivers.

Our studies with the practical transceiver design highlight that robust MCS for broadcast/multicast provide more flexibility to trade-off bit-rates (or coverage enhancements) between the two-layers carrying broadcast/multicast signals and unicast signals. The BMUST transceiver architecture has also shown significant improvements over OM in terms of increased SE or SNR gains (i.e. coverage enhancements) for the layer carrying unicast signals. Significant gains have also been verified with real channel estimation at the receivers where similar degradation is reported for OM and BMUST with imperfect interference cancellation in realistic mobile environments with very long delay spread and a simple channel estimation algorithm. These improvements can help fulfill some of the 5G industry requirements for broadcast/multicast [45] (cf. Section 9.1), such as efficient broadcast/multicast and unicast multiplexing and reuse of existing hard- and software capability for unicast to improve reliability and capacity, which motivate the potential inclusion of our considered two-layered superposition scheme into the evolution of the LTE-A-Pro and NR physical layers. Future work could consider extensions to MIMO spatial multiplexing and beamforming, multicell deployments, and other issues not considered in this work such as waveform design, system simulations with different radio resource scheduling algorithms and more sophisticated receiver architectures.

APPENDIX
PROOF OF PROPOSITION 2

One of the general properties of BC without feedback is that the BCs that have user channels with the same marginal probability distributions have the same capacity region [3]. This implies that the capacity region of the fading GBC satisfying (16) is unaffected by making the assumption:

$$\Phi_1^{(i)} = \Phi_2^{(i)} = \cdots = \Phi_K^{(i)} \equiv \Phi^{(i)}. \quad (21)$$

With this assumption, the received signal at one user may be written as the received signal at another user plus an independent noise. Hence, the channel is stochastically degraded and the rate region $\mathcal{C}_f$ in (15) derived with the assumption of channel degradedness becomes the capacity region. Moreover, the superposition coding achieves the capacity.

The weighted sum-rate optimality of the two-layer superposition in $\mathcal{C}_f$ can be then established by steps similar to the proof of Proposition 1 using the utility functions method. For convenience, we first rewrite the utility functions from (11) in terms of $\bar{Z}_k^{(i)}$ and $\Phi^{(i)}$ as

$$u_{\pi^{(i)}(k)}^{(i)}(\alpha) \equiv \frac{\mu/\mu_0}{\bar{Z}_{\pi^{(i)}(k)}^{(i)} \left|\Phi^{(i)}\right|^{-2} + \alpha} - \left(\frac{\lambda \ln 2}{\mu_0 \, N^{(i)}/N}\right); k = [1:K],$$

$$u_0^{(i)}(\alpha) \equiv \sum_{k=1}^{K} \frac{\mu_{0k}/\mu_0}{\bar{Z}_k^{(i)} \left|\Phi^{(i)}\right|^{-2} + \alpha} - \left(\frac{\lambda \ln 2}{\mu_0 \, N^{(i)}/N}\right).$$

The user orderings $\pi^{(i)}(1)$, $\pi^{(i)}(2)$, ..., $\pi^{(i)}(K)$ are now based on the long-term average noise powers only and not on the instantaneous channel realisations. To use the utility functions method as in the proof of Proposition 1, we need to check if the utility functions are still properly ordered on average. Due to $\bar{Z}_{\pi^{(i)}(1)}^{(i)} < \bar{Z}_{\pi^{(i)}(2)}^{(i)} < \cdots < \bar{Z}_{\pi^{(i)}(K)}^{(i)}$ and using the assumption (21), the utility functions corresponding to the independent messages satisfy

$$\mathbb{E}_{\Phi^{(i)}}\left(u_{\pi^{(i)}(1)}^{(i)}(\alpha)\right) > \mathbb{E}_{\Phi^{(i)}}\left(u_{\pi^{(i)}(k)}^{(i)}(\alpha)\right) \text{ for all } k \neq 1,$$

over all $\alpha > 0$. Therefore, the best user's message is selected over all the other users' messages and it is optimal to allocate all power to the best user's message *and* the common message in $i^{th}$ channel. In other words, for $\mu_0 > \mu$,

$$\max_{\mathbf{R} \in \mathcal{R}_f} \mu_0 R_0 + \mu \sum_{k=1}^{K} \sum_{i=1}^{M} R_k^{(i)} = \max_{\mathbf{R} \in \mathcal{R}_f} \mu_0 R_0 + \mu \sum_{i=1}^{M} R_{\pi^{(i)}(1)}^{(i)}.$$

This indicates that the weighted sum-rate point of $\mathcal{C}_f$ has zero rates for all $R_{\pi^{(i)}(2)}^{(i)}$, $R_{\pi^{(i)}(3)}^{(i)}$, …, $R_{\pi^{(i)}(K)}^{(i)}$, and 2-layer is optimal.

The resulting rate region is given by averaging individual rates in (7) over the channel statistics. It is easily seen that all boundary points of this rate region, including the weighted sum-rate point, are achievable using superposition coding and successive decoding. This completes the proof of Proposition 2. ∎

ACKNOWLEDGMENT

The authors would like to thank Jordi J. Gimenez and Simon Elliot for their valuable comments on realistic signal propagation models for the mean path loss based on the Okumura-Hata model, and the reviewers for their comments and suggestions that helped improve the paper.

REFERENCES

[1] M. Armstrong et al., "Object-Based Broadcasting – Curation, Responsiveness and User Experience, " BBC Research & Development, White Paper WHP 285, Sept. 2014.
[2] M. Evans et al., "Creating Object-Based Expriences in the Real World," in *Proc. Int. Broadcast. Convention* (*IBC*), Amsterdam, Sept. 2016.
[3] T. Cover and J. A. Thomas, *Elements of Information Theory*, Hoboken, NJ, USA:Wiley, 1991.
[4] 3GPP TS 36.201 v14.1.0, "Evolved Universal Terrestrial Radio Access (E-UTRA); LTE physical layer; General description", Sept. 2017.
[5] 3GPP TS 38.201 v15.0.0, "NR; Physical layer; General description", Sept. 2017.
[6] D. Tse, "Optimal Power Allocation over Parallel Gaussian Broadcast Channels," in *Proc. IEEE Int. Symp. Inform. Theory*, Ulm, Germany, p. 27, June 1997; Full version in https://web.stanford.edu/~dntse/papers/broadcast2.pdf.
[7] J. Jang and K. B. Lee, "Transmit Power Adaptation for Multiuser OFDM Systems," *IEEE J. Sel. Areas Commun.*, vol. 21, no. 2, pp. 171-178, Feb. 2003.
[8] A. El Gamal, "Capacity of the product and sum of two unmatched broadcast channels," *Probl. Information Transmission*, vol. 16, no. 1, pp. 3-23, Jan.-Mar. 1980.
[9] N. Jindal and A. Goldsmith, "Optimal Power Allocation for Parallel Gaussian Broadcast Channels with Independent and Common Information," in *Proc. IEEE Int. Symp. Inform. Theory*, Chicago, IL, pp. 215, June 2004.
[10] H. Weingarten, Y. Steinberg, and S. Shamai, "The Capacity Region of the Gaussian Multiple-Input Multiple-Output Broadcast Channel," *IEEE Trans. Inf. Theory*, vol. 52, no. 9, pp. 3936–3964, Sept. 2006.
[11] Y. Geng and C. Nair, "The Capacity Region of the Two-Receiver Gaussian Vector Broadcast Channel With Private and Common Messages," *IEEE Trans. Inform. Theory*, vol. 60, no. 4, pp. 2087-2104, Apr. 2014.
[12] Y. Saito et al., "Non-Orthogonal Multiple Access (NOMA) for Cellular Future Radio Access," in *Proc. IEEE Veh. Tech. Conf.-Spring*, Dresden, Germany, 5 pages, June 2013.
[13] Z. Ding, Z. Yang, P. Fan and H. V. Poor, "On the Performance of Non-Orthogonal Multiple Access in 5G Systems with Randomly Deployed Users," *IEEE Signal Process. Lett.*, vol. 21, no. 12, pp. 1501-1505, Dec. 2014.
[14] L. Dai et al., "Non-orthogonal multiple access for 5G: solutions, challenges, opportunities, and future research trends," *IEEE Commun. Mag.*, vol. 53, no. 9, pp. 74-81, Sept. 2015.
[15] B. Clerckx, H. Joudeh, C. Hao, M. Dai, and B. Rassouli, "Rate splitting for MIMO wireless networks: a promising PHY-layer strategy for LTE evolution," *IEEE Commun. Mag.*, vol. 54, no. 5, pp. 98-105, May 2016.
[16] Z. Ding et al., "Application of Non-Orthogonal Multiple Access in LTE and 5G Networks," *IEEE Commun. Mag.*, vol. 55, no. 2, pp. 185 – 191, Feb. 2017.
[17] Z. Ding et al. "A Survey on Non-Orthogonal Multiple Access for 5G Networks: Research Challenges and Future Trends," *IEEE J. Sel. Areas Commun.*, vol. 35, no. 10, pp. 2181-2195, Oct. 2017.
[18] Q. Sun, S. Han, C.-L. I, and Z. Pan, "On the Ergodic Capacity of MIMO NOMA Systems," *IEEE Wireless Commun. Lett.*, vol. 4, no. 4, pp. 405-408, Aug. 2015.

[19] Y. Yuan et al., "Non-Orthogonal Transmission Technology in LTE Evolution," *IEEE Commun. Mag.*, vol. 54, no. 7, pp. 68-74, July 2016.

[20] L. Zhang, et al., "Layered-Division-Multiplexing: Theory and Practice," *IEEE Trans. Broadcast.*, vol. 62, no. 1, pp. 216-232, Mar. 2016.

[21] J. Choi, "Minimum Power Multicast Beamforming With Superposition Coding for Multiresolution Broadcast and Application to NOMA Systems," *IEEE Trans. Commun.*, vol. 63, no. 3, pp. 791-800, Mar. 2015.

[22] S. I. Park et al., "Low Complexity Layered Division Multiplexing for ATSC 3.0," *IEEE Trans. Broadcast.*, vol. 62, no. 1, pp. 233-243, Mar. 2016.

[23] D. Gómez-Barquero and O. Simeone, "LDM Versus FDM/TDM for Unequal Error Protection in Terrestrial Broadcasting Systems: An Information-Theoretic View," *IEEE Trans. Broadcast.*, vol. 61, no. 4, pp. 571-579, Dec. 2015.

[24] D. Kim, F. Khan, C. V. Rensburg, Z. Pi, and S. Yoon, "Superposition of Broadcast and Unicast in Wireless Cellular Systems," *IEEE Commun. Mag.*, vol. 46, no. 7, pp. 110-117, July 2008.

[25] U. Sethakaset and S. Sun, "Sum-Rate Maximization in the Simultaneous Unicast and Multicast Services with Two Users," in *Proc. IEEE Int. Symp. Personal Indoor and Mobile Radio Commun.* (*PIMRC*), Istanbul, Turkey, Sept. 2010, pp. 672-677.

[26] L. Zhang et al., "Improving LTE eMBMS With Extended OFDM Parameters and Layered-Division-Multiplexing," *IEEE Trans. Broadcast.*, vol. 63, no. 1, pp. 32-47, Mar. 2017.

[27] Z. Ding, Z. Zhao, M. Peng, and H. V. Poor, "On the Spectral Efficiency and Security Enhancements of NOMA Assisted Multicast-Unicast Streaming," *IEEE Trans. Commun.*, vol. 65, no. 7, pp. 3151-3163, July 2017.

[28] O. Tervo et al., "Energy-Efficient Joint Unicast and Multicast Beamforming with Multi-Antenna User Terminals," in *Proc. IEEE Int. Workshop Sig. Proc. Adv. Wireless Commun.* (*SPAWC*), Sapporo, Japan, 5 pages, July 2017.

[29] Y.-F. Liu, C. Lu, M. Tao, and J. Wu, "Joint Multicast and Unicast Beamforming for the MISO Downlink Interference Channel," in *Proc. IEEE Int. Workshop Sig. Proc. Adv. Wireless Commun.* (*SPAWC*), Sapporo, Japan, pp. 159-163, July 2017.

[30] E. Chen, M. Tao and Y.-F. Liu, "Joint Base Station Clustering and Beamforming for Non-Orthogonal Multicast and Unicast Transmission With Backhaul Constraints," *IEEE Trans. Wireless Commun.*, vol. 17, no. 9, pp. 6265-6279, Sept. 2018.

[31] J. Zhao, O. Simeone, D. Gündüz, and D. Gómez-Barquero, "Non-Orthogonal Unicast and Broadcast Transmission via Joint Beamforming and LDM Cellular Networks," in *Proc. IEEE Global Commun. Conf.* (*GLOBECOM*), Washington, DC, 6 pages, Dec. 2016.

[32] E. G. Larsson and H. V. Poor, "Joint Beamforming and Broadcasting in Massive MIMO," *IEEE Trans. Wireless Commun.*, vol. 15, no.4, pp. 3058-3070, Apr. 2016.

[33] Y. Mao, B. Clerckx, and V. O. K. Li, "Rate-Splitting for Multi-Antenna Non-Orthogonal Unicast and Multicast Transmission," in *Proc. IEEE Int. Workshop Sig. Proc. Adv. Wireless Commun.* (*SPAWC*), Kalamata, Greece, 5 pages, June 2018.

[34] Y. Mao, B. Clerckx, and V. O. K. Li, "Rate-Splitting for Multi-Antenna Non-Orthogonal Unicast and Multicast Transmission: Spectral and Energy Efficiency Analysis," *arXiv preprint arXiv:1808.08325*, Aug. 2018; revised May 2019.

[35] 3GPP TS 36.211 v14.4.0, "Evolved Universal Terrestrial Radio Access (E-UTRA); Physical channels and modulation", Sept. 2017.

[36] Y. Liang, V. V. Veeravalli, and H. V. Poor, " Resource Allocation for Wireless Fading Relay Channels: Max-Min Solution," *IEEE Trans. Inform. Theory*, vol. 53, no. 10, pp. 3432-3453, Oct. 2007.

[37] D. Tunietti and S. Shamai (Shitz), "Fading Gaussian Broadcast Channels with State Information at the Receivers," in *Advances in Network Information Theorey*, P. Gupta, G. Kramer, and A. J. van Wijngaarden, Eds., DIMACS Series in Discrete Mathematics and Theoretical Computer Science, vol. 66, American Mathematical Society, pp.139-150, 2004.

[38] K. Marton, "A Coding Theorem for the Discrete Memoryless Broadcast Channel," *IEEE Trans. Inform. Theory*, vol. 25, no. 3, pp. 306-311, May 1979.

[39] S. Sadr, A. Anpalagan, and K. Raahemifar "Radio Resource Allocation for the Downlink of Multiuser OFDM Communication Systems," *IEEE Commun. Surveys & Tut.*, vol. 11, no.3, pp. 92-106, Third Quarter 2009.

[40] D. Vargas et al., "Practical Performance Measurements of LTE Broadcast (eMBMS) for TV Applications," in *Proc. Int. Broadcast. Convention* (*IBC*), Amsterdam, Sept. 2018.

[41] 3GPP TS 36.213 v14.4.0, "Evolved Universal Terrestrial Radio Access (E-UTRA); Physical layer procedures (Release 14)," Sept. 2017.

[42] 3GPP TS 36.212 v14.4.0, "Evolved Universal Terrestrial Radio Access (E-UTRA); Multiplexing and channel coding," Sept. 2017.

[43] 3GPP TR 38.901 v15.0.0, "Study on Channel Model for Frequencies from 0.5 to 100 GHz," June 2018.

[44] T. Stockhammer, I. Bouazizi, F. Gabin, J.M. Guyot, C. Lo, and T. Lohmar, "Enhanced TV Services over 3GPP MBMS," in *Proc. Int. Broadcast. Convention* (*IBC*), Amsterdam, Sept. 2017.

[45] 3GPP TS 38.913 v14.3.0, "Study on Scenarios and Requirements for Next Generation Acess Technologies", June 2017.

[46] 3GPP TR 36.776 v16.0.0, "Evolved Universal Terrestrial Radio Access (E-UTRA); Study on LTE-based 5G terrestrial broadcast", Mar. 2019.